\documentclass[aps,preprintnumbers,  amsmath, showpacs]{revtex4}
\usepackage{psfrag}
\usepackage{epsfig}
\usepackage{amsfonts} 
\usepackage{amssymb}  
\usepackage{graphicx} 
\usepackage{amsmath}  
\usepackage{amsmath}  
\usepackage{calligra}
\usepackage{mathrsfs}
\usepackage[latin1]{inputenc}
\usepackage{color}
\usepackage{suffix}
\usepackage{mathtools}
\usepackage{romannum}

\DeclareMathOperator{\sech}{sech}
\DeclareMathAlphabet{\mathcalligra}{T1}{calligra}{m}{n}
\DeclareFontShape{T1}{calligra}{m}{n}{<->s*[2.2]callig15}{}
\newcommand{\scriptr}{\mathcalligra{r}\,}

\newcommand{\bea}{\begin{eqnarray}}
\newcommand{\eea}{\end{eqnarray}}
\newcommand{\beas}{\begin{eqnarray*}}
\newcommand{\eeas}{\end{eqnarray*}}

\DeclarePairedDelimiterX\MeijerM[3]{\lparen}{\rparen}%
{\begin{smallmatrix}#1 \\ #2\end{smallmatrix}\delimsize\vert\,#3}

\newcommand\MeijerG[8][]{%
  G^{\,#2,#3}_{#4,#5}\MeijerM[#1]{#6}{#7}{#8}}

\WithSuffix\newcommand\MeijerG*[7]{%
  G^{\,#1,#2}_{#3,#4}\MeijerM*{#5}{#6}{#7}}

\def\Title#1{\begin{center} {\Large {\bf #1} } \end{center}}

\begin{document}

\Title{Thermodynamics of Neutrons in a Magnetic Field and its Implications for Neutron Stars}

\author{E. J. Ferrer and A. Hackebill}
\affiliation{Deptartment of Physics and Astronomy, CUNY College at Staten Island and CUNY Graduate Center, New York 10314, USA}
 
\begin{abstract}
We investigate the effects of a magnetic field on the thermodynamics of a neutron system at finite density and temperature. 
Our main motivation is to deepen the understanding of the physics of a class of neutron stars known as magnetars, which exhibit extremely strong magnetic fields.
Taking into account two facts, (i) the existence of a pressure anisotropy in the presence of a magnetic field and (ii) that the quantum field theory contribution to the pressure is non-negligible, we show that the maximum value that the inner magnetic field of a star can reach while being in agreement with the magnetohydrostatic equilibrium between the gravitational and matter pressures becomes $10^{17}$ G, an order of magnitude smaller than the previous value obtained through the scalar virial theorem; that the magnetic field has a negligible effect on the neutron system's equation of state; that the system's magnetic susceptibility increases with the temperature; and that the specific heat $C_V$ does not significantly change with the magnetic field in the range of temperatures characteristic of protoneutron stars.
\end{abstract}



\maketitle

\section{Introduction}

It is a well known fact that the global properties of neutron stars (NS) mostly depend on the equation of state (EOS) of their inner matter at densities above the saturation density $\rho_s = 2.8 \times 10^{14}$ g cm$^3$ (which corresponds to a nucleon density of $n_s \approx 0.16$ fm$^{-3}$). Close to those densities, nuclear star matter has a very small proton fraction, resulting in almost pure neutron matter (see  Refs. \cite{Chamel}-\cite{Lattimer-2016} for review). The expected density in the cores of massive neutron stars could be as large as $ \approx (5-10) \rho_s$.
In those highly dense cores, the neutron-rich matter can give rise to more exotic degrees of freedom, like hyperons \cite{Hyperons}, and perhaps even to a quark matter phase transition (see Ref. \cite{Lattimer2010} for review), hence forming a so-called hybrid star. Moreover, since cold strange quark matter could be absolutely stable \cite{Witten}, a phase transition may occur that would favor the creation of a quark-matter phase, thus giving rise to a strange star.

Despite an extended observational and theoretical effort carried out in recent years to determine the EOS of nuclear-star matter, there still remains many questions to be answered. The physics of matter at densities beyond $\rho_s$ is a big challenge to theorists, with observations of NS being crucial for determining the correct dense-matter model. In this regard, recent very precise mass measurements of two compact objects, PSR J1614-2230 and PSR J0348+0432 with $M=1.908\pm 0.016M_{\odot}$ \cite{Demorest} and $M=2.01\pm 0.04M_{\odot}$ \cite{Antoniadis}, respectively, where $M_{\odot}$ is the solar mass, have provided some constraints on the interior composition of NS. These high mass values imply that the EOS of the corresponding stellar medium should be rather stiff at high densities.
On the other hand,  the recent detection of gravitational waves produced by the binary NS merger event GW170817 \cite{Merger} will be crucial for shedding new light on the internal structure of NS. Even more, binary neutron star mergers could be a common source, with an expected detection rate of $ \approx 40 $ yr$^{-1}$, as predicted by population-synthesis models \cite{Abadie}. This opens the possibility for new advances in the determination of the compact-star EOS in the near future. 

We should emphasize that NS, apart from being the most compact observed objects in the universe, also exhibit the strongest magnetic fields known in nature. Some radio pulsars are endowed with surface magnetic fields of order $10^{13} - 10^{14}$ G \cite{Pulsar-B}. From spectroscopic and spin-down studies of soft-$\gamma$-ray repeaters (SGRs) and anomalous x-ray pulsars (AXPs), it has been inferred that surface magnetic fields of order $10^{14} - 10^{15}$ G occur in some special compact objects called magnetars \cite{Magnetars}. Moreover, the inner core magnetic fields of magnetars can be even larger as follows from the magnetic field flux conservation in stellar media with very large electric conductivities. The inner fields have been estimated to range from $10^{18}$ G for nuclear matter stars \cite{Nuclear-matter-field} to $10^{20}$ G for quark matter stars \cite{Quark-Matter-field}. The fact that strong magnetic fields populate the vast majority of astrophysical compact objects and that they have significant consequences for several star properties has motivated a lot of work focused on the study of the EOS of magnetized neutron stars, both without considering  \cite{Quark-Matter-field}-\cite{Chaichian} and considering \cite{EoS-AMM}-\cite{Insignificance} the magnetic-field interaction with the particle anomalous-magnetic-moment. 

An important characteristic of the stellar medium EOS in a uniform magnetic field is that they become anisotropic, with different pressures in directions along and transverse to the field respectively \cite{Quark-Matter-field, Canuto}. In Ref. \cite{Quark-Matter-field}, the transverse and longitudinal pressures were found by taking the quantum-statistical average of the energy-momentum tensor using the path-integral approach. The obtained results coincide with those obtained years ago by using the many-particle density matrix \cite{Canuto}. In this regard, there exists a controversy about whether the stellar medium should be considered as a fluid satisfying Pascal's law. We will discuss this point in more detail at the end of Sec. \Romannum{3}.

In calculating the EOS in a magnetic field, the contribution of charged particles will be affected by the Landau quantization of their transverse momenta and by the interaction between the magnetic field with the particles' anomalous magnetic moments (B-AMM); while for neutral composite particles, such as neutrons, only the second contribution is present. The effect of the B-AMM interaction in the EOS of charged particles was recently proved in Ref. \cite{Insignificance} to be insignificantly small both in the strong-field and the weak-field approximations. For neutrons, however, it was found in Ref. \cite{Lattimer} that for critical fields of order $\approx 10^{18}$ G the magnetic field can make a substantial contribution to the system EOS. 

In this paper, we want to revise this result by taking into account the effect of the anisotropy in the  EOS of a magnetized system, as well as the effect of the quantum field theory (QFT) contribution, that as we will show in Sec. \Romannum{2}A, plays an important role for this effective model. We will show that while the transverse pressure increases with the magnetic field, the parallel pressure decreases, reaching zero value at a magnetic field strength, which is on the order of the one needed to produce a significant effect under the isotropic neutron system EOS assumption. 

Motivated by the fact that the longitudinal pressure vanishes at sufficiently high magnetic fields, we make an alternative investigation to determine the maximum value that the inner magnetic field of a neutron star formed by neutrons can have when the matter pressure and the star's gravitational pressure are in magneto-hydrostatic equilibrium, taking into account the matter pressure anisotropy in the presence of a magnetic field. The value of the maximum inner magnetic field obtained by this method is on the order of $10^{17}$ G, an order of magnitude smaller than the one obtained by considering the equipartition of the gravitational and magnetic energies in the framework of the scalar virial theorem \cite{Nuclear-matter-field}. This result indicates that for the allowed magnetic-field range, the effect of the B-AMM interaction has a negligibly small effect, similar to the case of charged particles studied in Ref. \cite{Insignificance}. 
 
Finally, we investigate the thermal behavior of the magnetized neutron system. We find a peculiar temperature dependence of the system's magnetic susceptibility: It increases with temperature in the range of temperatures of interest for proto-neutron stars. This unusual behavior coincides with the one recently found by lattice QCD calculations for the deconfined quark-gluon plasma in a magnetic field \cite{Bonati}. We also find that the specific heat in the temperature range of interest for proto-neutron stars does not have a significant magnetic-field dependence.

The paper is organized as follows. In Sec. \Romannum{2}, we introduce the effective model of a neutron system in the presence of a magnetic field and calculate the different components of the corresponding thermodynamic potential in the one-loop approximation. In Sec. \Romannum{3}, we investigate the effect of the magnetic field on the system's EOS, while discussing the important roles that the quantum field theory contribution and the anisotropy of the pressures can play in this effective model. In Sec. \Romannum{4}, we investigate the temperature and density dependence of the system's magnetic susceptibility, as well as how it can be affected by the magnetic field. In Sec. \Romannum{5}, the dependence of the system's specific heat on temperature and the magnetic field is studied. Sec. \Romannum{6} is devoted to estimating the maximum value the inner magnetic field can reach in a NS when the matter and gravitational pressures are in magneto-hydrostatic equilibrium according to a Newtonian framework. Finally, in Sec. \Romannum{7}, we summarize the main results of this investigation.

\section{Neutron many-particle system in a magnetic field}\label{section2}

If the inner density of a compact star is not sufficiently high to produce the deconfinement of quarks, the nuclear star matter below the crust will be formed mainly by neutrons.  With this context in mind, we wish to investigate the effects of a magnetic field on a neutron system. We consider an effective theory in which the neutrons, although electrically neutral, interact with a  magnetic field via their anomalous magnetic moment (AMM). The magnetic field is taken to be uniform in the $\hat{z}$ direction. The effective Lagrangian density is then given by

\begin{equation}
\label{Lagrangian}
    \mathscr{L}=\bar\Psi_N(i\gamma_{\mu}{\partial}^\mu-M_N+i{k_N}{\sigma_{\mu\nu}F^{\mu\nu})\Psi_N},
\end{equation}
with $k_N$ representing the neutron's anomalous magnetic moment given by $k_N=\mu_N g_N/2$, where $\mu_N=|e|\hbar/2M_P=3.15\times 10^{-18}$ MeV/G  is the nuclear magneton  \cite{Lande-Factor}, $M_P$ is the proton mass and $g_N=-3.82$ is the Land$\acute{e}$ $\textit{g}$ factor for the neutron, whose negative sign indicates that the neutron's magnetic moment is similar to that of a negatively charged particle even though the neutron is a composite neutral particle. In (\ref{Lagrangian}), $M_N=939.56$ MeV is the neutron mass, $ \sigma_{\mu\nu}= \frac{i}{2}[\gamma_\mu,\gamma_\nu]$, and $F^{\mu\nu}$ is the electromagnetic field strength tensor corresponding to the applied uniform and constant magnetic field.

The parameter $k_N$, which is playing the role of the coupling constant for the B-AMM interaction, has dimensions of inverse energy ($k_N \approx 1/E$). Thus, the dimensionless parameter, $k_N E$, is small at low energies and large at high energies. This means that the corresponding theory is non-renormalizable and we have to consider it as an effective theory that is reliable only up to a certain energy scale that is defined by the leading parameter in the system under consideration. In the case of a neutron star, it is natural to consider the baryonic chemical potential as the physical scale.

The dispersion relation in momentum space corresponding to (\ref{Lagrangian}) is given by
\begin{equation}
\label{Dispersion-Eq}
   \det(-{\gamma}^{\mu}{p_\mu}-M_N-i{k_N}B{\gamma_2}{\gamma_1})=0,
\end{equation}
where we took the external magnetic field to be in the $z$ direction (i.e., $B=F^{21}$). From the solution of the dispersion relation (\ref{Dispersion-Eq}), we obtain the energy spectrum 
\begin{equation}
\label{Energy-Spectrum}
    E_{\eta,\sigma}={\eta}\sqrt{{p_3}^2+\Big(\sqrt{{M_N}^2+{p_1}^2+{p_2}^2}+{\sigma}{k_N}B\Big)^2}
\end{equation}
Here $\eta=\pm$ denotes particle-antiparticle modes and $\sigma=\pm$ denotes spin-up and spin-down modes. As expected, the B-AMM interaction breaks the spin degeneracy that exists at zero magnetic field.

To study the many-particle theory, we calculate the grand-canonical thermodynamic potential in the one-loop approximation,
\begin{equation}
\label{Thermodynamic-Potential}
    \Omega_N(B,\mu,T)={\frac{-1}{\beta} \sum_{p_4}{\int{\frac{d^3p}{(2\pi)^3}\ln[\det(-{\gamma}^{\mu}{p^*_\mu}-M_N-i{k_N}B{\gamma_2}{\gamma_1})]}}},
\end{equation}
where $\beta=1/T$ denotes the inverse of the absolute temperature, and the baryonic chemical potential $\mu$ appears as a shift in the Euclidean four-momentum ($p_0=ip_4$, with  $p_4=\frac{(2n+1)\pi}{\beta}, \quad n=0,\pm1,\pm2,...$,) in $p^*_\mu=(ip_4-\mu, p_1,p_2,p_3)$.

After taking the determinant in (\ref{Thermodynamic-Potential}), the thermodynamic potential can be given in terms of the energy spectrum (\ref{Energy-Spectrum}) as
\begin{equation}
\label{Omega-2}
    \Omega_N(B,\mu,T)=\frac{-1}{\beta} \sum_{p_4}\sum_{\sigma=\pm} \int{\frac{d^3p}{(2\pi)^3} \ln\left[(p_4+i\mu)^2+(E_{+,\sigma})^2\right]}.
\end{equation}

Performing the sum in Matsubara frequencies we obtain
\begin{equation}\label{Omega-N}
    \Omega_N=\Omega_{QFT}+\Omega_{S}+\Omega_0,
\end{equation}
where $\Omega_{QFT}$ is the quantum-field-theory contribution (the one that does not depend on the chemical potential and temperature) given by
\begin{equation}
\label{Vac-Cont}
    \Omega_{QFT}=-\sum_{\sigma=\pm}\int_{-\infty}^{\infty}{\frac{d^3p}{(2\pi)^3}(E_{+,\sigma})},
\end{equation}
$\Omega_S$ is the statistical contribution, which depends on the system temperature and chemical potential
\begin{equation} \label{Omega-beta}
    \Omega_{S}=-\frac{1}{\beta}\sum_{\sigma=\pm}\int_{-\infty}^{\infty}{\frac{d^3p}{(2\pi)^3}\Big(\ln[1+e^{-{\beta}(E_{+,\sigma}+\mu)}]+\ln[1+e^{-{\beta}(E_{+,\sigma}-\mu)}]\Big)},
\end{equation}
and $\Omega_0$ is a term added to make the vacuum pressure equal to zero when all the physical parameters, $\mu$, $T$, etc., are zero, as well as to guarantee the renormalization of the QFT contribution at $B=0$.
From here, we will proceed by analyzing the different contributions separately.

\subsection{QFT contribution}

Starting with the QFT term (\ref{Vac-Cont}), we can write it by switching to cylindrical coordinates ($p_1=p_{\perp}\cos{\theta}$, $p_2=p_{\perp}\sin{\theta}$) as
\begin{equation}
    \Omega_{QFT}=\frac{-1}{2\pi^2}\sum_{\sigma}\int_{0}^{\infty}p_{\perp}dp_\perp\int_{0}^{\infty}dp_3\sqrt{p_3^2+\Big[\sqrt{M_N^2+p_{\perp}^2}+{\sigma}k_NB\Big]^2}.
\end{equation}

The integral diverges, but as we discussed earlier, the theory under consideration is an effective theory that is only valid up to a certain energy scale. Thus,
we introduce the energy cutoff $\Lambda$, such that
\begin{equation}
    \Lambda^2\ge{p_3^2+\Big(\sqrt{M_N^2+p_{\perp}^2}+{\sigma}k_NB\Big)^2}.
\end{equation}
As we indicated above, this energy scale should be presumably fixed in a neutron star by the baryonic chemical potential (i.e., $\Lambda \approx \mu$).

On this region, we have
\begin{equation}
    \Omega_{QFT}=\frac{-1}{2\pi^2}\sum_{\sigma}\int_{0}^{p_{\perp}^{Max}}p_{\perp}dp_\perp\int_{0}^{p_{3}^{Max}}dp_3\sqrt{p_3^2+\Big[\sqrt{M_N^2+p_{\perp}^2}+{\sigma}k_NB\Big]^2},
\end{equation}

where
\begin{equation}
    p_{3}^{Max}=\sqrt{\Lambda^2-\big[\sqrt{M_N^2+p_\perp^2}+\sigma{k_N}B\big]^2},\quad p_{\perp}^{Max}=\sqrt{(\Lambda-\sigma{k_N}B)^2-M_N^2}.
\end{equation}

Letting $p_3'=\frac{p_3}{\Lambda}$, $p_\perp'=\frac{p_\perp}{\Lambda}$, $\tilde{p}_{3}^{Max}=\frac{p_{3}^{Max}}{\Lambda}$, $\tilde{p}_{\perp}^{Max}=\frac{p_{\perp}^{Max}}{\Lambda}$, $\tilde{M}_N=\frac{M_N}{\Lambda}$, and $\tilde{B}=\frac{{\sigma}{k_N}{B}}{\Lambda}$, $\Omega_{QFT}$ can be expressed after integrating in $p_3'$ as
\begin{equation}
   \nonumber{ \Omega_{QFT}=\frac{-\Lambda^4}{2\pi^2}\sum_\sigma\int_{0}^{\tilde{p}_{\perp}^{Max}}dp_\perp'\frac{p_\perp'}{2}\Bigg[\sqrt{1-\big(\tilde{B}+\sqrt{\tilde{M}_N^2+p_\perp'^2}\big)^2}}
\end{equation}

\begin{equation}
    \nonumber{-\big(\tilde{B}+\sqrt{\tilde{M}_N^2+p_\perp'^2}\big)^2\ln\bigg[\tilde{B}+\sqrt{\tilde{M}_N^2+p_\perp'^2}\bigg]}
\end{equation}

\begin{equation}
    +\big(\tilde{B}+\sqrt{\tilde{M}_N^2+p_\perp'^2}\big)^2\ln\bigg[1+\sqrt{1-\big(\tilde{B}+\sqrt{\tilde{M}_N^2+p_\perp'^2}\big)^2}\bigg]\Bigg].
\end{equation}

After integrating in $p_\perp'$ and performing the sum in $\sigma$ this becomes

\begin{eqnarray} \label{Magnetization-1}
	  &\Omega_{QFT}=\frac{-1}{48{\pi}^2}\Bigg(\Lambda^2\bigg[\sqrt{\frac{\Lambda^2-(M_N+Bk_N)^2}{\Lambda^2}}((M_N-Bk_N)^2-4M_N^2+6\Lambda^2)+(B \to -B)\bigg]
\nonumber
\\
 &+8k_N\Lambda^3\Big[B\sin^{-1}\Big[\frac{Bk_N-M_N}{\Lambda}\Big]+(B \to -B)\Big]
\nonumber
\\
 &+B^4k_N^4\bigg[\ln\Big[1+\sqrt{\frac{\Lambda^2-(M_N+Bk_N)^2}{\Lambda^2}}\Big]-\ln\Big[\frac{M_N-Bk_N}{\Lambda}\Big]+(B \to -B)\bigg]
\nonumber
\\
  &+M_N^2\bigg[(6B^2k_N^2-8Bk_NM_N+3M_N^2)\ln\Big[\frac{M_N-Bk_N}{\Lambda}\Big]+(B \to -B)\bigg]
 \nonumber
\\
   &-M_N^2\bigg[(6B^2k_N^2+8Bk_NM_N+3M_N^2)\ln\Big[1+\sqrt{\frac{\Lambda^2-(M_N+Bk_N)^2}{\Lambda^2}}\Big]+(B \to -B)\bigg]\Bigg).
\end{eqnarray}
The introduced notation $(B \to -B)$ stands for the same contribution except with the sign of the magnetic field flipped. The existence of the two terms with different magnetic field signs is associated to the spin-up and spin-down 
contributions in the magnetic-moment and magnetic-field interaction.

 Taylor expanding in $\frac{M_N}{\Lambda}$, $\frac{k_NB}{\Lambda}$, we find
\begin{eqnarray}\label{Omega-vac}
{\Omega_{QFT}\simeq\frac{-1}{4\pi^2}\Big[\Lambda^4+(k_N^2B^2-M_N^2){\Lambda^2}+\Big(\frac{M_N^4}{2}+M_N^2k_N^2B^2-\frac{k_N^4B^4}{6}\Big)\ln\Big[\frac{M_N}{2\Lambda}\Big]}
\nonumber
\\
+\Big(\frac{M_N^4}{8}+\frac{3}{2}M_N^2k_N^2B^2-\frac{k_N^4B^4}{6}\Big)+\mathcal{O}(\frac{1}{\Lambda})\Big].\quad\quad\quad\quad\quad\quad
\end{eqnarray}

From this result, we have that the first three terms in the right-hand side of (\ref{Omega-vac}) diverge for $\Lambda \rightarrow \infty$. Moreover, in the starting Lagrangian, there are no appropriate bare terms to absorb those terms depending on $\kappa_N B$ through a suitable renormalization procedure, as should be expected for this nonrenormalizible theory. This is corroborating that our model for neutrons in a magnetic field is an effective model that is only valid at energies below $\Lambda \approx \mu$.

From (\ref{Omega-vac}), we see that the first term on its right-hand side does not depend on the physical parameters, but only on the cutoff $\Lambda$. Thus, to guarantee that the vacuum pressure vanishes appropriately in vacuum, we need to take out this term in Eq. (\ref{Omega-N}). Moreover, we need to regain the renormalized thermodynamic potential for B=0. Thus, to satisfy both requirements, we improve the regularization procedure proposed in Ref. \cite{Universe} by taking
\begin{equation}\label{Zero-vacuum}
   \Omega_0=-\Omega_{QFT}(B=0, \Lambda),
\end{equation}
which means subtracting in Eq. (\ref{Omega-N}) the QFT contribution that depends on $\Lambda$ at $B=0$.

On the other hand, the term $k_N^2B^2{\Lambda^2}$ in Eq. (\ref{Omega-vac}) makes a large field contribution to the thermodynamic potential for $\Lambda \approx \mu$
and it should be considered on equal footing with the many-particle contribution when calculating the field-dependent EOS.
We will see as follows that the QFT contribution will have a non-negligible impact on the system EOS, something that has been ignored in previous works.

\subsection{Many-particle contribution}

Since the baryonic chemical potential $\mu$ is expected to be the leading parameter in the stellar medium, we expect $\Omega_{S}$ to make a substantial contribution on the system EOS. Moreover, we take into account that due to the cooling effect produced by neutrino emission during the protoneutron star epoch,  $T\ll \mu$ for a relatively old neutron star. Hence, we concentrate our calculation now on the $T=0$ case. Then, letting $\Omega_{\mu}=\lim_{{\beta}\to\infty}\Omega_{S}$, we have from (\ref{Omega-beta}) that
\begin{equation} \label{Omega-mu}
    \Omega_{\mu}=-\int_{-\infty}^{\infty}{\frac{d^3p}{(2\pi)^3}\big[(\mu-E_{+,-}){\Theta}(\mu-E_{+,-})+(\mu-E_{+,+}){\Theta}(\mu-E_{+,+})\big]},
\end{equation}
which after integration gives
\begin{eqnarray} \label{Omega-mu-2}
    \Omega_{\mu}&=&\frac{-1}{48{\pi}^2}\Bigg[2\bigg(\sqrt{1-\Big({\frac{M_N+{k_N}B}{\mu}}\Big)^2}+(B \to -B)\bigg){\mu^4} +4{k_N}B\bigg({\sin^{-1}}\Big(\frac{M_N+{K_N}B}{\mu}\Big)-(B \to -B)\bigg){\mu^3}
  \nonumber    
\\
    &+&\Bigg[(M_N+{k_N}B)^3(3M_N-{k_N}B)\Bigg(\ln\bigg[1+\sqrt{1-\Big({\frac{M_N+{k_N}B}{\mu}}\Big)^2}\bigg]-\ln\Big[\frac{\big|M_N+{k_N}B\big|}{\big|\mu\big|}\Big]\Bigg)+(B \to -B)\Bigg] 
 \nonumber  
\\
     &+&\Bigg [(8{k_N}B(M_N+{k_N}B)-5(M_N+{k_N}B)^2)\sqrt{1-\Big({\frac{M_N+{k_N}B}{\mu}}\Big)^2}+(B \to -B)\Bigg]{\mu^2}\Bigg].
\end{eqnarray}

From (\ref{Omega-mu-2}), we see that in order for the potential to remain real the arguments of the radical terms must be non-negative, which implies 
\begin{equation}
    \Big|\frac{M_N\pm{k_N}B}{\mu}\Big|\leq1.
\end{equation}

The chemical potential must therefore always be greater than or equal to the given combination of the neutron mass and ${k_N}B$ term (i.e., $\mu\ge1$GeV). Furthermore, for a given value of the chemical potential, the magnitude of the magnetic field is bounded by 
\begin{equation} \label{B-max}
    B\leqslant{\Big|\frac{\mu-M_N}{k_N} \Big|}.
\end{equation}

Hence, for values of $\mu$ near $2.1$ GeV, which is the maximum value for NS obtained by assuming causality and the validity of the isotopic TOV equations \cite{Causality},
the upper bound of the magnetic field strength is approximately $B_{max}(\mu=2.1$ GeV$) \approx1.93\times10^{20}$ G. We call attention to the fact that this is not a valid approximation because, as we will see, at those magnetic-field values the pressure splitting is significant, which is indicating that the isotopic TOV equations are not a good approximation.

\section{Equation of State of a many-particle neutron system in a magnetic field}\label{section3}

The EOS of many-particle systems in the presence of a magnetic field has been previously studied by several authors \cite{Lattimer}, \cite{EoS-AMM}-\cite{Insignificance}. Since neutrons in a magnetic field do not suffer the softening in the pressure caused by Landau quantization it has been found that the magnetic field, through its interaction with the neutron magnetic moment, stiffens the system EOS if the pressure anisotropy is not considered and the Maxwell pressure is disregarded \cite{Lattimer}. In this section we will show that due to the pressure anisotropy produced by the magnetic field \cite{Quark-Matter-field}-\cite{ Canuto}, the situation is more subtle in the sense that at the magnetic-field strength needed to produce a significant stiffening of the EOS the longitudinal pressure vanishes creating a loss of balance with the gravitational pull in the neutron star.  

Let us review now the physical basis for the anisotropic structure of the EOS in a magnetic field. The origin of the anisotropy in the pressures is connected with the breaking of the rotational SO(2) symmetry by the uniform magnetic field. In a covariant formulation \cite{Quark-Matter-field}, the external uniform magnetic field has an associated normalized electromagnetic strength tersor $\widehat{F}^{\mu \rho}=F^{\mu \rho} /B$, which can split the Minkowski metric into two structures, $\eta_{\perp}^{\mu \nu}=\widehat{F}^{\mu \rho}\widehat{F}_{\rho}^\nu$ and $\eta_{\|}^{\mu \nu}=\eta^{\mu \nu}-\widehat{F}^{\mu \rho}\widehat{F}_{\rho}^\nu$. Hence, under these conditions, the quantum-statistical average of the energy-momentum tensor of the magnetized many-particle system can be divided into the following three covariant structures,
\begin{equation}\label{Tau}
\frac{1}{\beta V}\langle{\tau}^{\mu \nu}\rangle=\Omega_N \eta^{\mu \nu}+(\mu N+TS)u^\mu u^\nu +BM \eta_{\perp}^{\mu \nu},
\end{equation}
where $N$ is the particle number density, $S$ is the entropy, $M$ is the system magnetization, $V$ is the system volume, $\beta=1/T$ is the inverse absolute temperature, and $u_\mu$ is the medium four-velocity, which in the rest system takes the value $u_\mu =(1,\overrightarrow{0})$. In vacuum, only the first term on the right-hand side of (\ref{Tau}) is present. Once statistics (temperature and/or density) are in place, the Lorentz symmetry is broken and the second term on the right-hand side of  (\ref{Tau}) acknowledges it through the presence of the four-velocity of the medium, $u_\mu$; finally, the last term on the right hand side of  (\ref{Tau}) gives rise to the breaking of the rotational symmetry through the new tensor $\eta_{\perp}^{\mu \nu}$, which takes place in the presence of a magnetic field.

Thus, the presence of a uniform magnetic field generates an anisotropy in the energy-momentum tensor, such that the pressure parallel to the direction of the magnetic field $(P_\parallel)$ differs from the pressure perpendicular to the field direction $(P_\perp)$ \cite{Quark-Matter-field}-\cite{ Canuto}. For a magnetic field along the third-spatial direction, the energy density $ \varepsilon$ and pressures, $P_\parallel$ and  $P_\perp$, are given by 
\begin{equation} \label{energy-pressure}
    \varepsilon=\frac{1}{{\beta}V}\big<{\tau}^{00}\big>,\quad P_\parallel=\frac{1}{{\beta}V}\big<{\tau}^{33}\big>,\quad P_\perp=\frac{1}{{\beta}V}\big<{\tau}^{\perp\perp}\big>.
\end{equation}
Here $\big<{\tau}^{\rho\lambda}\big>$ is the quantum-statistical average of the energy momentum tensor given by 
\begin{equation} \label{energy-momentum-tensor}
    \big<{\tau}^{\rho\lambda}\big>=\frac{Tr[{\tau}^{\rho\lambda}e^{-\beta(H-{\mu}N)}]}{Z},
\end{equation}
where, $H$ is the system Hamiltonian and 
\begin{equation} \label{energy-momentum-tensor-2}
    {\tau}^{\rho\lambda}=\int_{0}^{\beta}d\tau\int{d^3x[\tau^{\rho\lambda}_M+\tau^{\rho\lambda}_N]}
\end{equation}
with $\tau^{\rho\lambda}_M$and $\tau^{\rho\lambda}_N$ being the contributions to the energy momentum tensor arising from the pure magnetic field (Maxwell contribution) and from the many-particle neutron system respectively, and $Z$ being the partition function of the grand-canonical ensemble, which is given by
\begin{equation} \label{partition function}
    Z=Tr[e^{-\beta(H-{\mu}N)}].
\end{equation}

The energy density and pressures in (\ref{energy-pressure}) can be expressed in terms of the thermodynamic potential $\Omega_N$  \cite{Quark-Matter-field}. Considering the $T\to0$ limit, (\ref{energy-pressure}) becomes
\begin{equation} \label{energy-density}
    \varepsilon=\Omega_N^{(T=0)}-\mu\frac{\partial{\Omega_\mu}}{\partial{\mu}},
\end{equation}

\begin{equation} \label{Pressures}
    P_\parallel=-\Omega_N^{(T=0)}-\frac{B^2}{2},\quad{P_\perp=-\Omega_N^{(T=0)}-MB+\frac{B^2}{2}}
\end{equation}
where the quadratic terms in $B$ arise from the Maxwell contribution to the energy momentum tensor.  

The system magnetization in (\ref{Pressures}) is given by
\begin{equation} \label{Magnetization}
   M=-\frac{\partial\Omega_N^{(T=0)}}{\partial{B}},
\end{equation}
which results in 
\begin{eqnarray} \label{Magnetization-2}
	&M=\frac{k_N}{12\pi^2}\Bigg(\bigg[2M_N\bigg[\sqrt{1-\frac{(M_N-Bk_N)^2}{\Lambda^2}}-(B \to -B)\bigg] +Bk_N\bigg[\sqrt{1-\frac{(M_N-Bk_N)^2}{\Lambda^2}}+(B \to -B)\bigg]\bigg]\Lambda^2
\nonumber
\\
 &+(2\Lambda^3+\mu^3)\bigg[\sin^{-1}\bigg[\frac{Bk_N+M_N}{\Lambda}\bigg]- (B \to -B)\bigg]
\nonumber
\\
 &+\Bigg[(2M_N+Bk_N)(M_N-Bk_N)^2\ln\bigg[1+\sqrt{1-\frac{(M_N-Bk_N)^2}{\Lambda^2}}\bigg]-(B \to -B) \Bigg]
\nonumber
\\
  &+\Bigg[(2M_N-Bk_N)(Bk_N+M_N)^2\ln\bigg[\frac{(M_N+Bk_N)^2}{\Lambda^2}\bigg]-(B \to -B) \Bigg]\Bigg) 
 \nonumber
\\
  &+\frac{k_N}{24\pi^2}\Bigg(2\mu^2\bigg[(M_N+2Bk_N)\sqrt{1-\frac{(M_N-Bk_N)^2}{\mu^2}}-(B \to -B)\bigg]
\nonumber	
\\
   &+\bigg[2(2M_N-Bk_N)(M_N+Bk_N)^2\ln\bigg[1+\sqrt{1-\frac{(M_N+Bk_N)^2}{\mu^2}}\bigg]-(B \to -B)\bigg]
\nonumber
\\
  &+\Bigg[(2M_N+Bk_N)(M_N-Bk_N)^2\ln\bigg[\frac{(M_N-Bk_N)^2}{\mu^2}\bigg]- (B \to -B) \Bigg]\Bigg).
\end{eqnarray}

We can see that the magnetization of this system gets a significant contribution from the magnetized vacuum, which is reflected in the $\Lambda$ dependence in Eq. (\ref{Magnetization-2}). This is a consequence of the appearance in the thermodynamic potential of terms that combine the magnetic field with the energy scale $\Lambda$. Other situations where the energy scale makes significant contributions in magnetized systems can be found in the value of the chiral condensate of the Nambu-Jona-Lasinio version of QCD \cite{Chiral-Cond} and in the gap of the magnetic color-flavor-locking phase of color superconductivity \cite{CS}. All this indicates that in finding the system magnetization the particle contribution alone will not be enough.
\begin{figure}
    \centering
    \includegraphics[scale=.25]{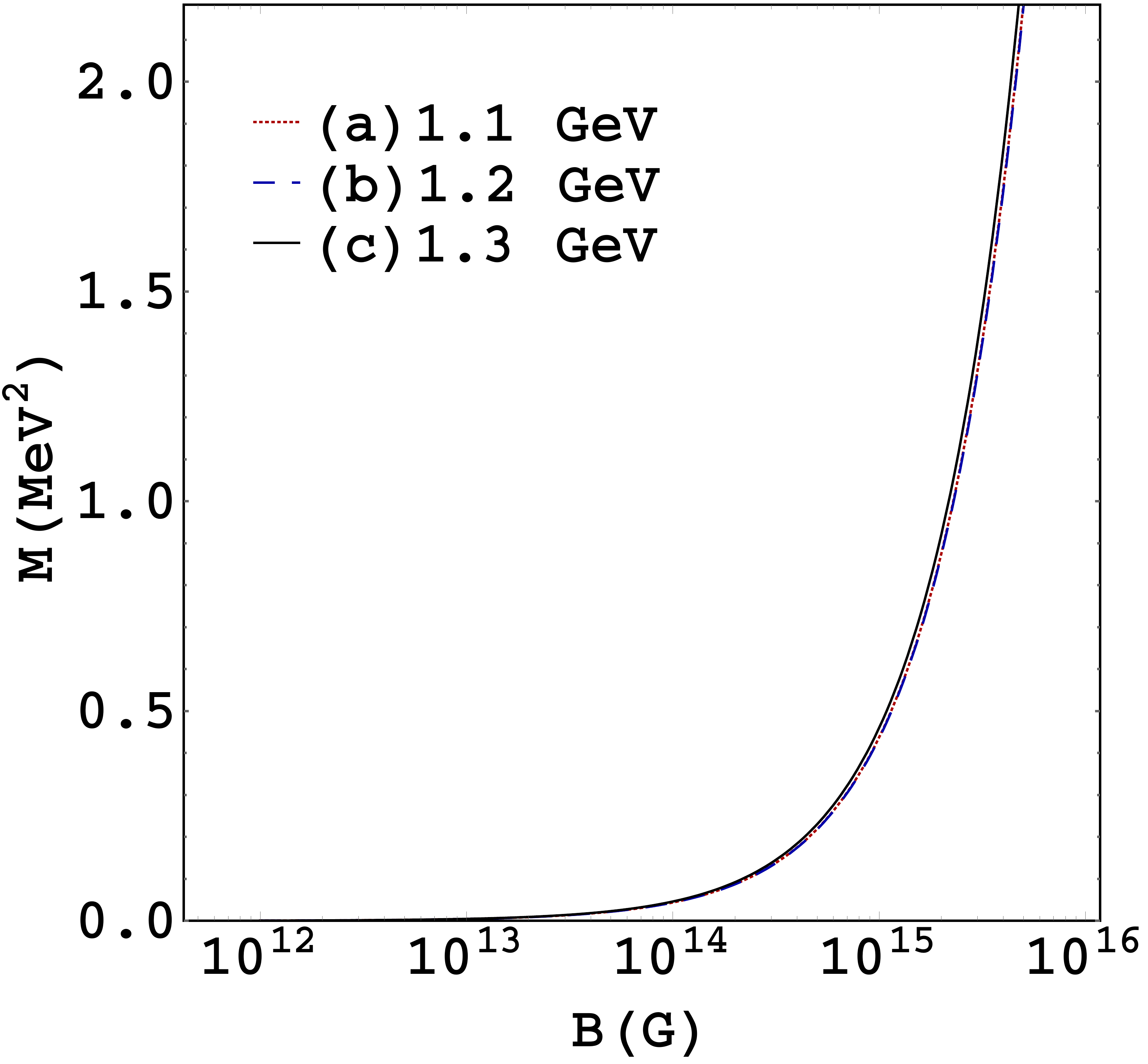}
    \caption{(Color online) Magnetization as a function of the magnetic field intensity at a fixed chemical potential. Curves with (a) dotted line, (b) discontinuous segments, and (c) continuous line have baryonic chemical potentials $\mu=1.3$ GeV,  $\mu=1.2$ GeV, and  $\mu=1.1$ GeV respectively. }
    \label{M-B}
\end{figure}
The magnetization versus the magnetic field for different values of the baryon density is plotted in Fig. \ref{M-B}. Notice there, that the magnetization monotonically increases with the magnetic field, which corresponds to a paramagnetic medium formed by neutral particles that do not exhibit the de Haas-van Alphen oscillations associated with Landau quantization \cite{Insignificance}. This result is expected since the field dependence of $M$ originates in the neutron magnetic moment interaction with the magnetic field, which is typically the dominant interaction in paramagnetism. Moreover, fixing the magnetic field, we can observe in these plots that  the magnetization increases with $\mu$. This is a consequence of the fact that neutrons and antineutrons have magnetic moments with different signs. Thus, having a larger positive baryonic chemical potential means there are more neutrons than antineutrons and so the net magnetic moment will be larger, hence producing a more noticeable magnetization.

\begin{figure}
\begin{center}
\includegraphics[width=0.47\textwidth]{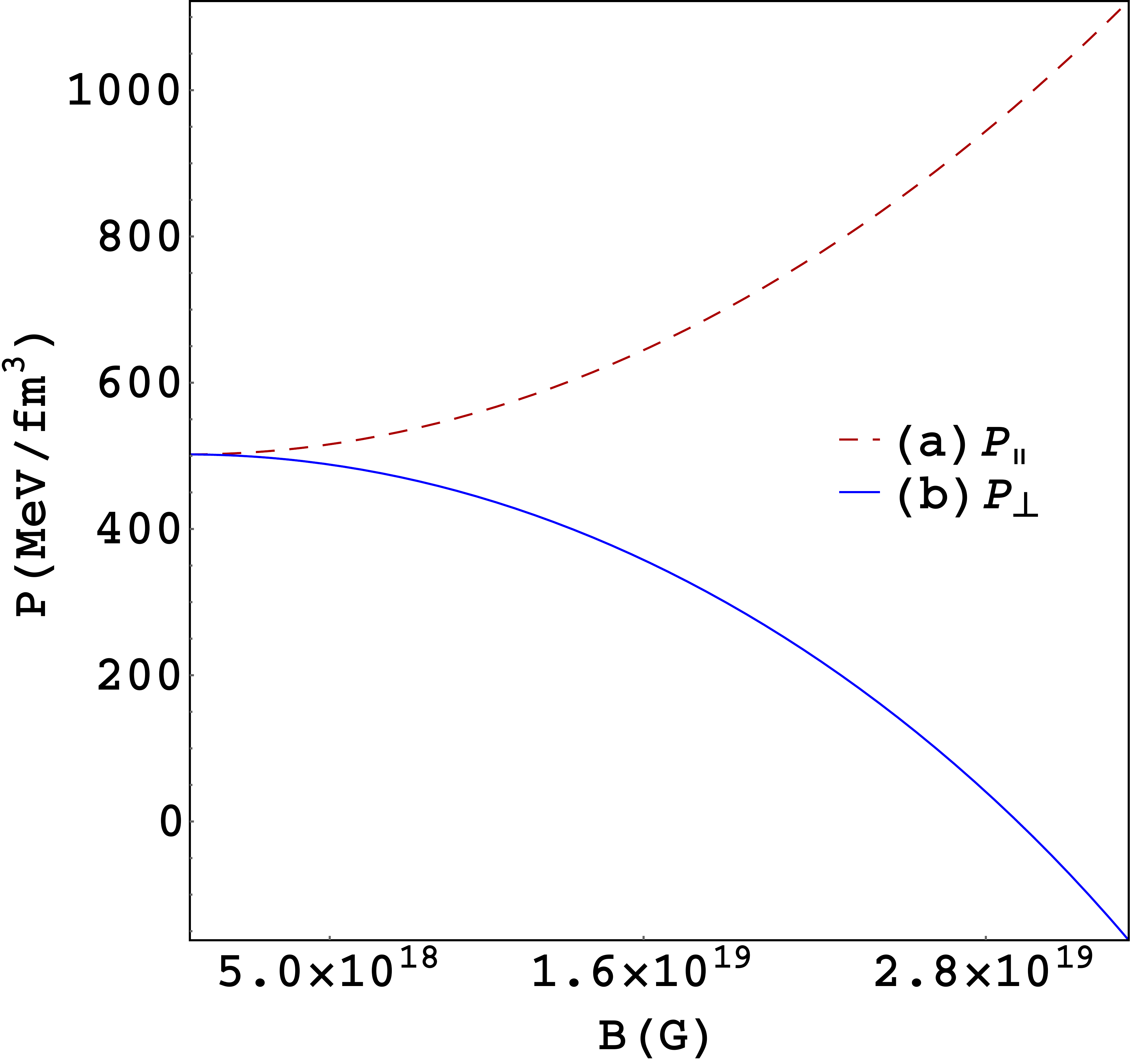}
\includegraphics[width=0.47\textwidth]{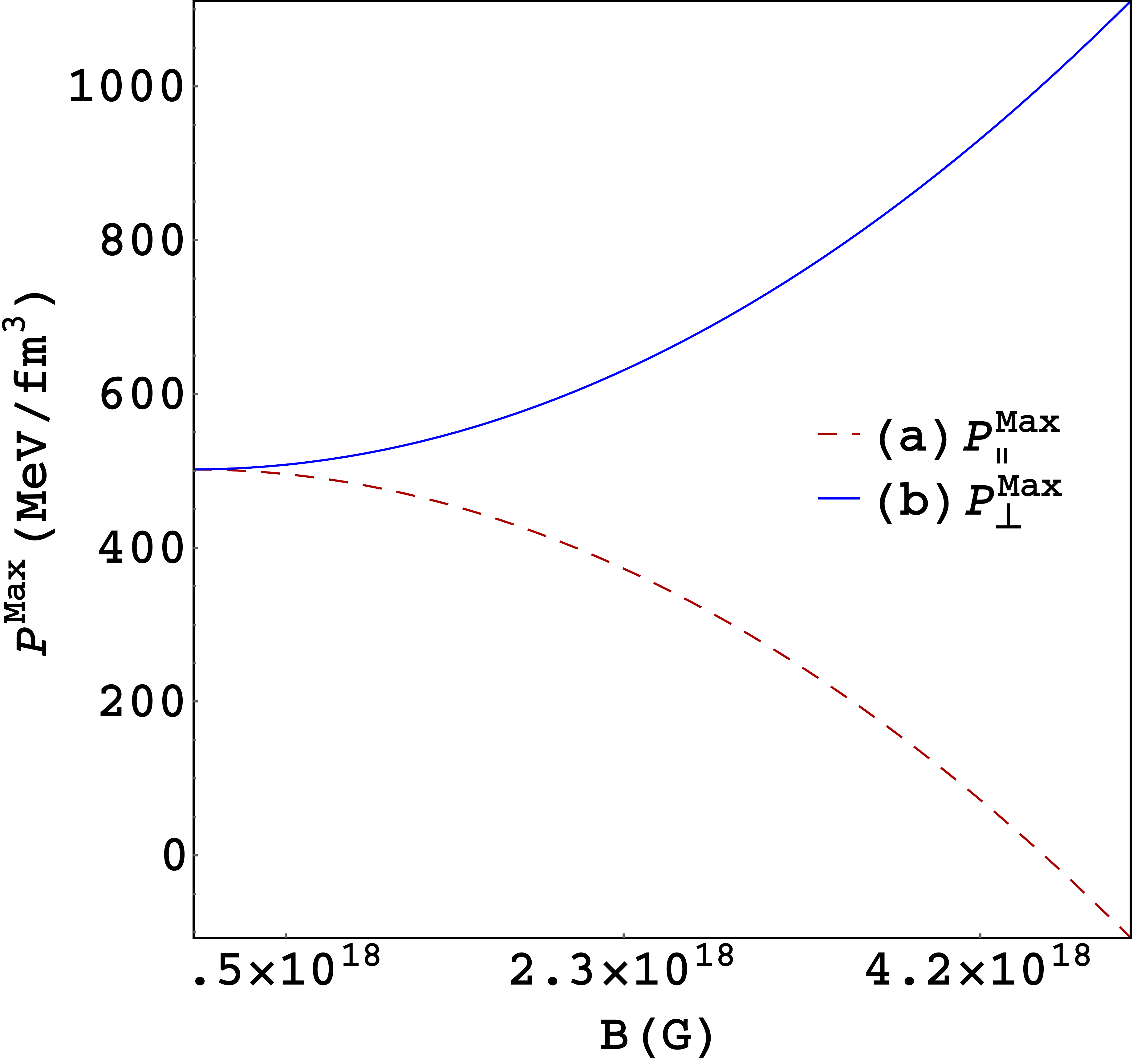}
\caption{(Color online) (a) Parallel and (b) perpendicular pressures without Maxwell contribution (left panel) and with Maxwell contribution (right panel), as a function of the magnetic field at a fixed chemical potential  $\mu=1.2$ GeV.}
\label{P-Splitting}
\end{center}
\end{figure}

In the left panel of Fig. \ref{P-Splitting}, the plots of the parallel and perpendicular pressures as obtained from (\ref{Pressures}) but without their Maxwell contributions, are presented as functions of $B$ for a chemical potential $\mu=1.2$ GeV. The chemical potential value $\mu=1.2$ GeV is needed to produce positive pressures at small fields. The contribution of the QFT thermodynamic potential depletes the pressures making necessary a higher Fermi pressure to produce a net positive pressure that eventually can compensate the gravitational pull.  
The perpendicular pressure decreases monotonically and has a root at $B^{(0)}_{\perp}(\mu=1.2$ GeV$)\approx2.9\times10^{19}$ G, which is lower than the bound field $B_{max}$($\mu=1.2$ GeV) $\approx 4.3\times10^{19}$ G obtained from Eq. (\ref{B-max}). In the right panel, the same plots are given but now with the inclusion of the Maxwell contributions from Eq. (\ref{Pressures}). In this case, while the perpendicular pressure increases with the magnetic field, the parallel pressure decays monotonically due to the sign of its Maxwell term and has a root at $B^{(0)}_\parallel$($\mu=1.2$ GeV)$\approx4.5\times10^{18}$ G, which is one order smaller than $B_{max}$($\mu=1.2$ GeV). 

We can see from comparing the results of the left and right panels of Fig. \ref{P-Splitting} that for the magnetic-field strengths under consideration, the Maxwell contributions to the pressures are dominant, since they can flip the pressures' behaviors with the magnetic field. Thus, at the field values where the splitting between the parallel and perpendicular pressures become significant, the pure magnetic pressures, coming from the Maxwell contributions,  determine the magnetohydrostatic equilibrium that should balance the gravitational inward  pull in NS's.

\begin{figure}
\begin{center}
\includegraphics[width=0.5\textwidth]{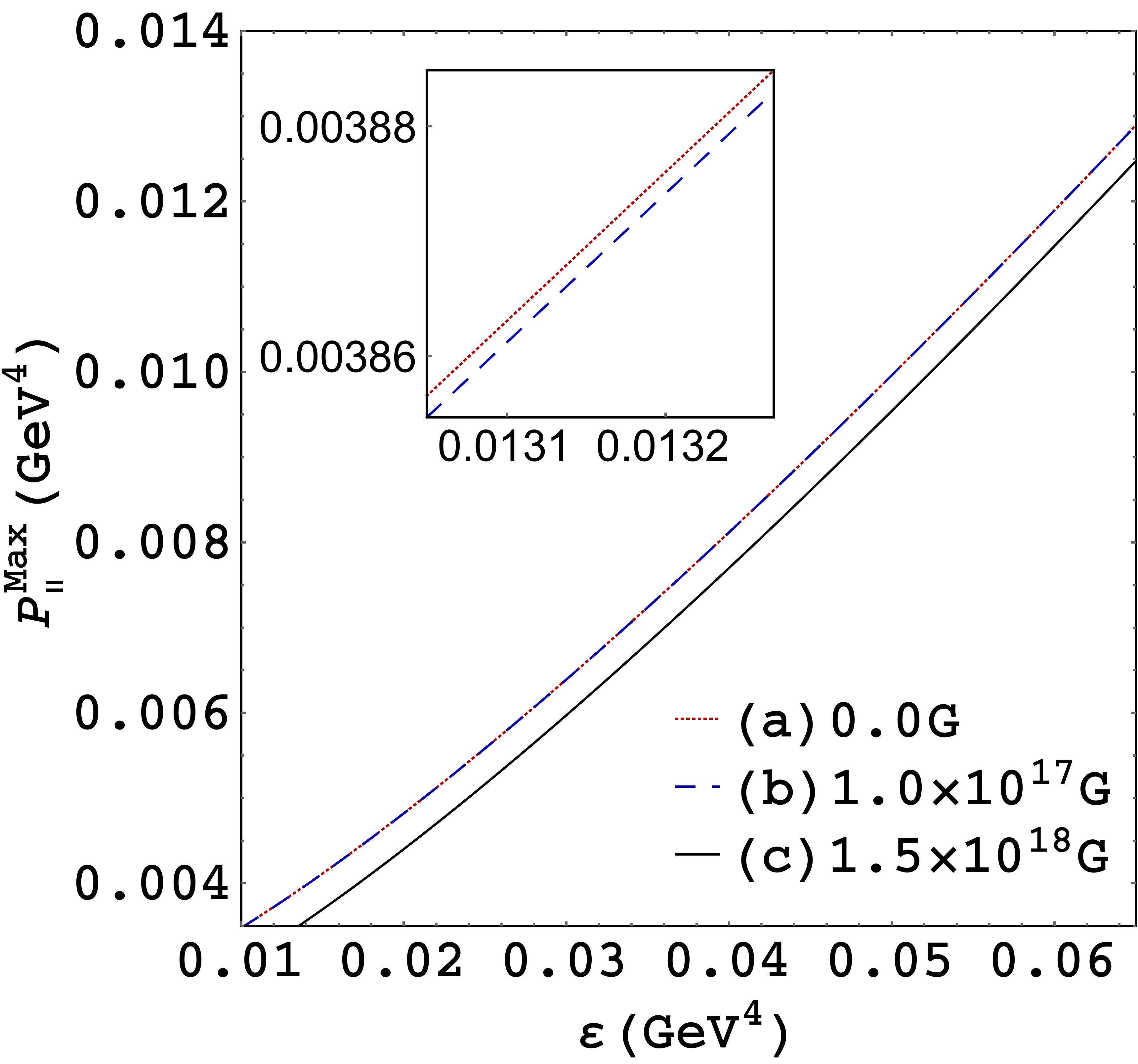}
\includegraphics[width=0.5\textwidth]{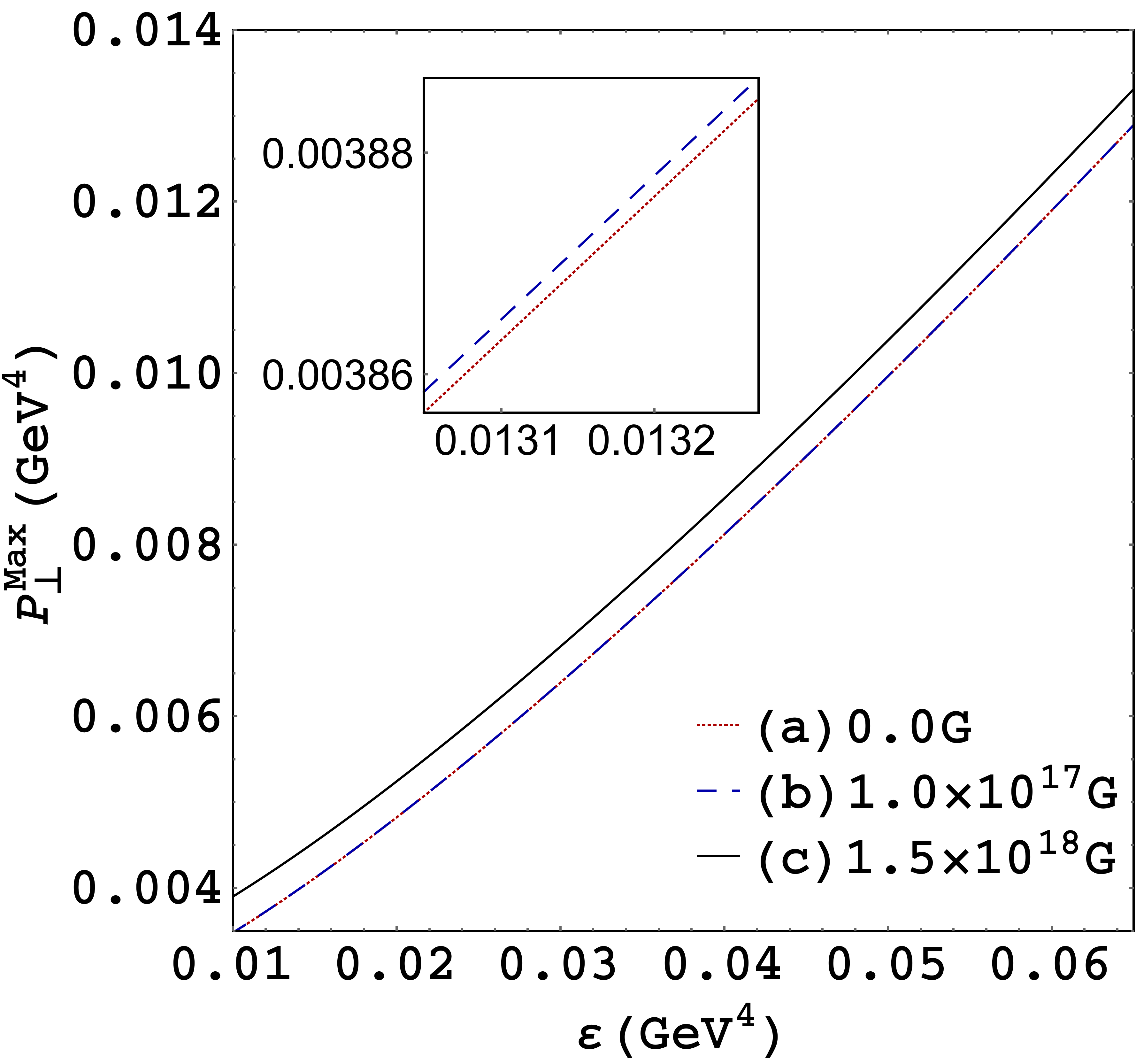}
\caption{(Color online) Equations of state for the parallel pressure with Maxwell contribution (upper panel) and perpendicular pressure with Maxwell contribution (lower panel), for magnetic field strengths of (a) $0.0G$ (dotted), (b) $1.0\times10^{17}G$ (dashed), and (c) $1.5\times10^{18}G$ (solid). The pressures at zero field and at $B \approx 10^{17}$ G overlap. (see in the inserted windows enlarged images for those field values).}
\label{EOS-Maxwell}
\end{center}
\end{figure}

 In Fig.  \ref{EOS-Maxwell}, the EOS's for the parallel (left panel) and perpendicular (right panel) pressures are plotted with their Maxwell contributions included. Here, the high field value of $B=1.5\times10^{18}$ G is chosen for its proximity to $B^{(0)}_{\parallel}$($1.2$ GeV).  As can be seen from both figures, the EOS's are not shifted significantly when the magnetic field strength increases from $0.0$ G to $1.0\times10^{17}$ G; however, there is a noticeable change when the magnetic field approaches $B^{(0)}_{\parallel}(1.2$ GeV). Another visible effect is that while a magnetic field of order $B \approx10^{18}$ G vertically shifts the EOS of the perpendicular pressure upward, it shifts the EOS of the parallel pressure downward.  Inside neutron stars, we expect the matter pressure to be positive. This suggests that physical pressures occur for $B\leq{B^{(0)}_{\parallel}}$ in systems where the Maxwell contribution is taken into consideration.
 Thus, the magnetic-field-strength values that could produce a significant shift in the EOS may be not physically realizable. 
  

 In Sec. \Romannum{6}, we will reconsider the problem of the maximum inner field that is tolerable to balance the gravitational pressure, by taking into account the anisotropy in the neutron degeneracy pressure due to the presence of the magnetic field, as well as the contribution of $\Omega_{QFT}$ in the matter pressure.
 
We should notice from Fig. \ref{EOS-Maxwell} that the magnetic field has a double effect on the system EOS. While for the perpendicular pressure component, the magnetic field helps to produce a vertical shift in the EOS, for the parallel component the opposite effect takes place. This is why the anisotropy generated by the magnetic field does not give rise to a unique effect in the slope of the EOS. Hence, to determine how the magnetic field will affect the star mass-radius relationship, the TOV equations need to be modified using a metric in agreement with the pressure anisotropy, which of course cannot be the spherical one considered in the original derivation of those equations \cite{TOV}.

\begin{figure}
\begin{center}
\includegraphics[width=0.47\textwidth]{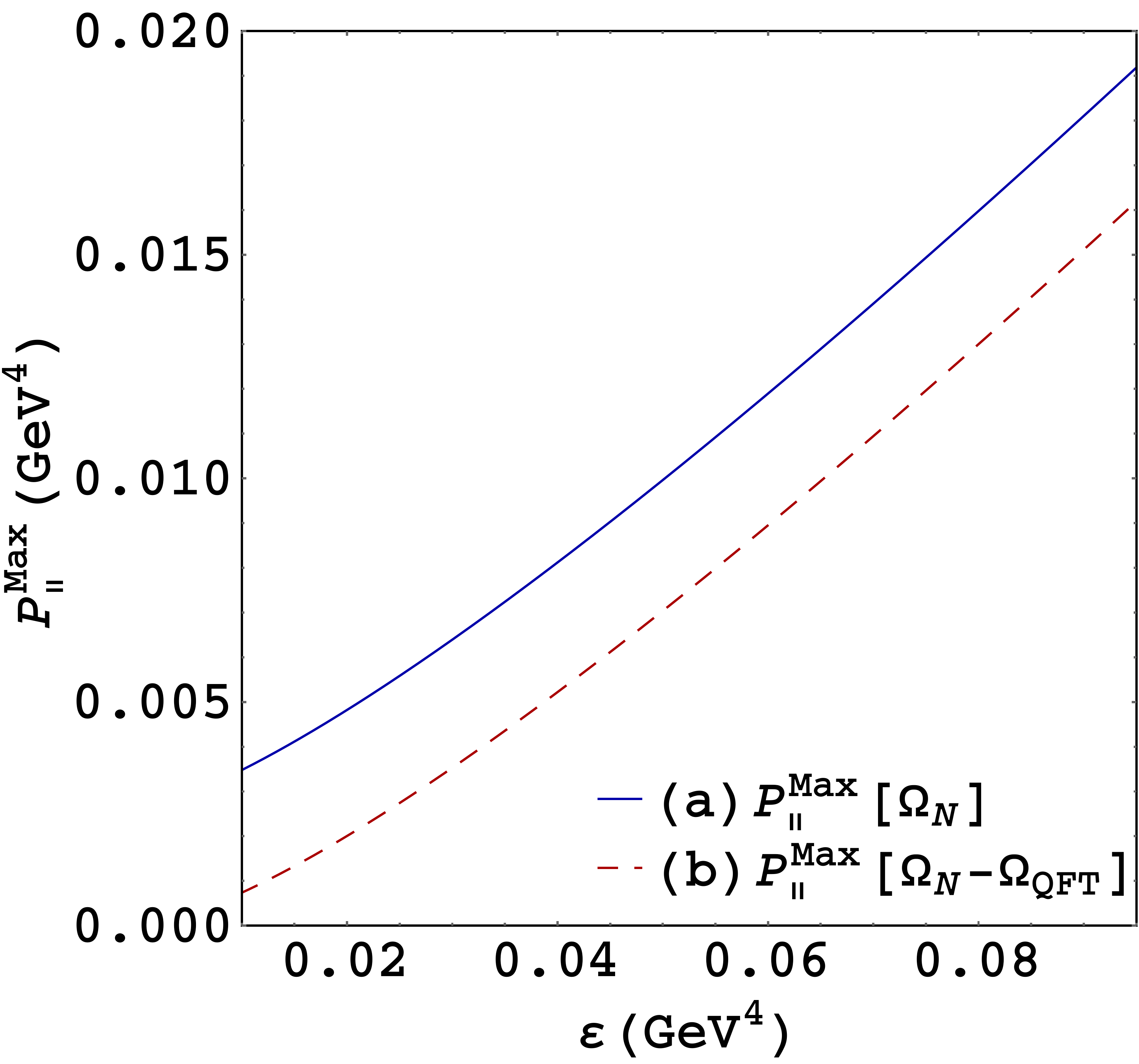}
\includegraphics[width=0.47\textwidth]{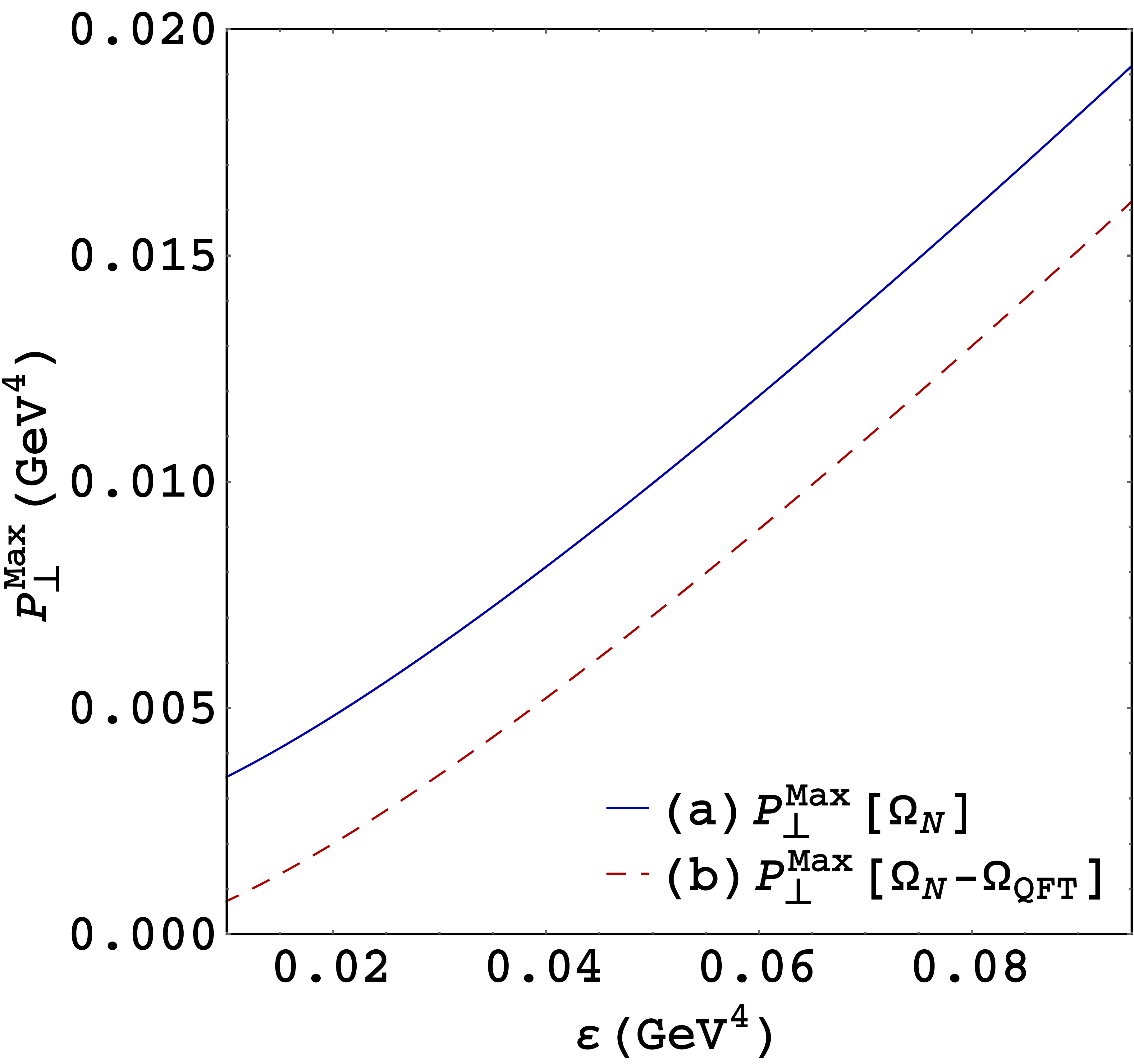}
\caption{(Color online) Equations of state for the parallel pressure with Maxwell contribution  (left panel) and for the perpendicular pressure with Maxwell contribution (right panel), (a) considering the pressure arising from the full thermodynamic potential $\Omega_N$ (continuous line) and (b) neglecting the $\Omega_{QFT}$ contribution (dashed line). The magnetic field under consideration is $B=10^{16}G$.}
\label{EOS-Comparison}
\end{center}
\end{figure}

Also, we should call attention to the fact that the $\Omega_{QFT}$ term makes a noticeable contribution to the system EOS, as can be seen in Fig. \ref{EOS-Comparison}. There, we notice that the inclusion of $\Omega_{QFT}$ significantly stiffens the EOS.

Finally, the following comment is in order. Regarding the anisotropy in the pressures, there exists a controversy. It was originally conjectured in Ref. \cite{Blandford} that the magnetization term in Eq. (\ref{Pressures}) is canceled by the Lorentz force induced by the magnetization. In that way, the matter pressure becomes isotropic. Some discussion against and in favor of this conjecture took place more recently in Refs. \cite{Potekhin, Oertel} and Refs. \cite{Vivian, Veronica}, respectively. In Ref. \cite{Oertel}, it was shown, for example, that in the energy-momentum conservation equation the contribution of the magnetization is canceled out by a term coming from the momentum of the so-called perfect-fluid contribution, but no demonstration was done in this paper, or in any other up to now, that demonstrates the cancellation of the magnetization term in the pressure. What is more, it is important to emphasize that even if the magnetization term is canceled, the pure Maxwell anisotropic contribution to the pressure will remain. Notwithstanding, several works on magnetized EOS's have decided to only consider the isotropic thermodynamic pressure (see, for example, Ref. \cite{Isotropic-Pressure}). As is known, the thermodynamic pressure $p$ is the only pressure that can exist in an isotropic fluid, where the spatial diagonal components of the energy-momentum tensor are all equal to $-p$. In this case, and only then, any rotation of the coordinate system not only keeps invariant the trace of the spatial-component of the energy-momentum tensor (i.e., the stress tensor), but also maintains the same value for each diagonal term. This is a well-known case where Pascal's law is satisfied, rendering the same pressure in any direction (due to the invariance of the diagonal components of the stress tensor) inside the fluid. In a magnetic field, for a coordinate system with axes formed along and perpendicular to the field, the stress tensor is diagonal but not isotropic; hence, we cannot define the thermodynamic pressure as the unique pressure of the system, and we must consider two pressures, one along and the other transverse to the field direction. If there exists a coordinate system where the stress tensor is diagonal but not isotropic, then Pascal's law is not satisfied. These arguments lead us to believe that it is fundamental to consider the pressure anisotropy in the study of magnetized NSs.

\section{Magnetic Susceptibility}\label{M-S}

 The magnetic susceptibility, $\chi_M$, is defined as the coefficient of the linear expansion of the magnetization in powers of the field, 
\begin{equation}\label{Magnetization-3}
	M=\chi_MB
\end{equation}

Knowing the system magnetic susceptibility we can know how strong the magnetization induced by a weak field can be (notice that in this approximation $\chi_M$ is constant, but for a sufficiently high magnetic field, $\chi_M$ can also depend on the field).

From (\ref{Magnetization})-(\ref{Magnetization-3}), we have that at zero temperature the magnetic susceptibility of the neutron system is given by
\begin{eqnarray}\label{Susceptibility-1}
\chi_{M_0}=\frac{k_N^2}{2\pi^2}\Bigg[\frac{M_N^2+\Lambda^2}{\sqrt{1-\frac{M_N^2}{\Lambda^2}}}+M_N^2\bigg[-\ln\bigg[1+\sqrt{1-\frac{M_N^2}{\Lambda^2}}\bigg]+\ln\Big[\frac{M_N}{\Lambda}\Big]\bigg]\Bigg]
 \nonumber
\\
+\frac{k_N^2}{4\pi^2}\Bigg[2\sqrt{1-\frac{M_N^2}{\mu^2}}\mu^2+M_N^2\bigg[2\ln\Big[1+\sqrt{1-\frac{M_N^2}{\mu^2}}\Big]-2\ln[\frac{M_N}{\mu}]\bigg]\Bigg].
\end{eqnarray}

If we take the cutoff $\Lambda$ to be equal to the chemical potential, then this further simplifies to

\begin{equation}\label{Susceptibility-2}
  	\chi_{M_0}=\frac{k_N^2\mu^3}{\pi^2\sqrt{\mu^2-M_N^2}}.
\end{equation}

\begin{figure}
    \centering
    \includegraphics[width=0.47\textwidth]{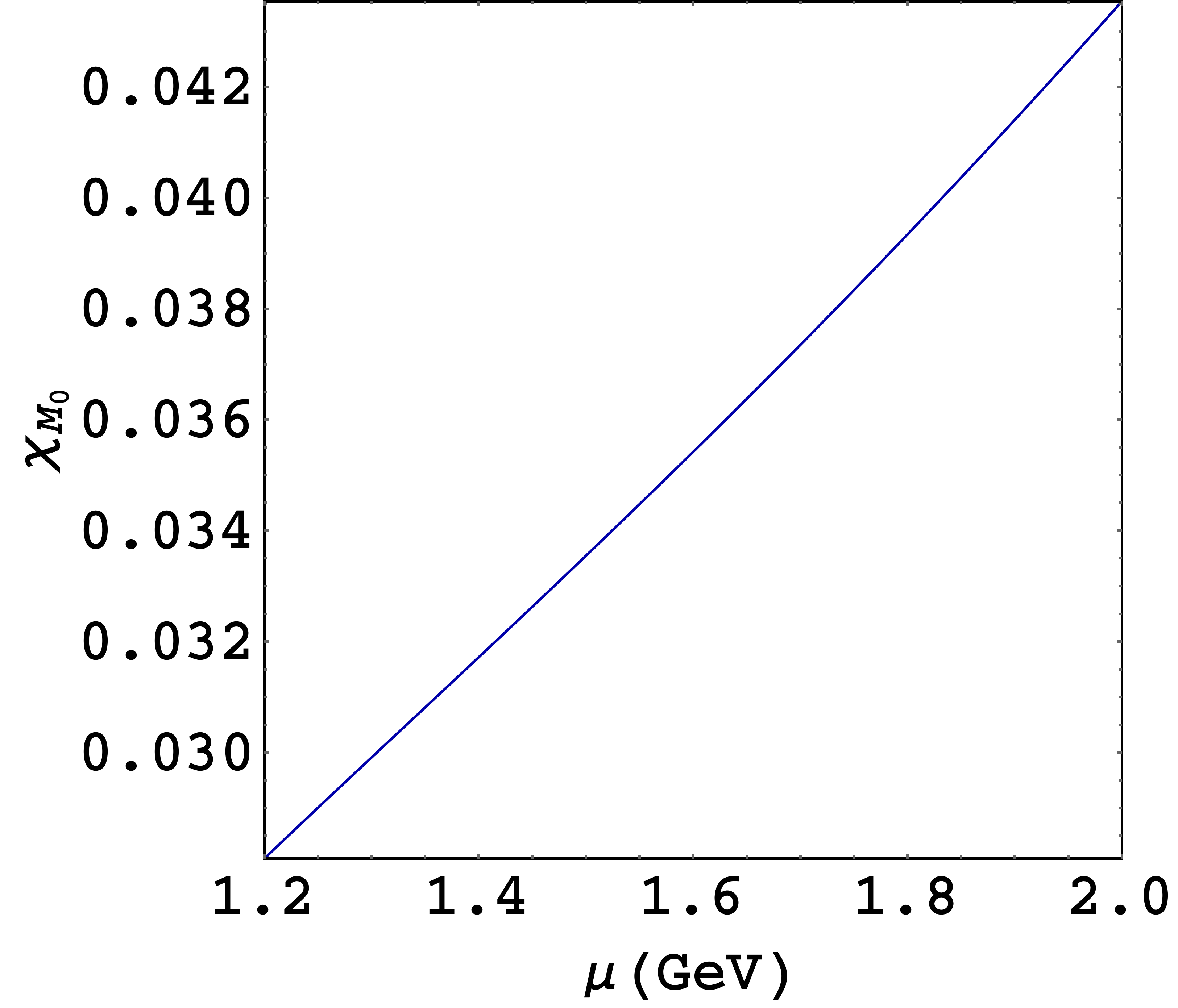}
    \caption{Magnetic susceptibility vs baryonic chemical potential at zero temperature.}
    \label{Sus-mu}
\end{figure}

In Fig. \ref{Sus-mu} we plot (\ref{Susceptibility-1}) versus the baryonic chemical potential. We can see how the susceptibility increases with $\mu$, in agreement with (\ref{Susceptibility-2}). This result is expected since the magnetization is also increasing with $\mu$ as we already showed. From a physical point of view, this can be understood from the fact that at zero temperature the system only has particles contributing to the total magnetization through their magnetic moments. When we increase the chemical potential, $\mu$, the radius of the Fermi sphere  increases, which implies an increase in the number of particles participating. Thus, the system's total magnetic moment, given as the sum of the individual magnetic moments of the neutrons, which are closed to the Fermi surface, increases.

To find how the temperature affects the magnetic susceptibility, we first have to calculate the temperature-dependent magnetization from (\ref{Omega-beta}), which is given as
\begin{equation}
	M_\beta=-\frac{\partial \Omega_S}{\partial B},
\end{equation}
whose linear term in the magnetic field is given by
\begin{equation}
	M_\beta\simeq\frac{T}{(2\pi)^3}\sum_{\sigma=\pm}\int_{-\infty}^{\infty}d^3p\Bigg[\frac{e^{\frac{\mu}{T}}k_N^2\Big[-Tp_3^2e^{\frac{\mu}{T}}-Tp_3^2e^{\frac{E(0)}{T}}+(E(0)^2-p_3^2)E(0)e^{\frac{E(0)}{T}}\Big]}{T^2E(0)^3\Big[e^{\frac{\mu}{T}}+e^{\frac{E(0)}{T}}\Big]^2}
	\nonumber
\end{equation}

\begin{equation}
    +\frac{k_N^2\Big[-Tp_3^2-Tp_3^2e^{\frac{1}{T}(\mu+E(0))}+(E(0)^2-p_3^2)E(0)e^{\frac{1}{T}(\mu+E(0))}\Big]}{T^2E(0)^3\Big[1+e^{\frac{1}{T}(\mu+E(0))}\Big]^2}\Bigg]B.
\end{equation}
where $E(0)=E_{+,\sigma}(B=0)=\sqrt{p_1^2+p_2^2+p_3^2+M_N^2}$ .

\begin{figure}
\begin{center}
\includegraphics[width=0.49\textwidth]{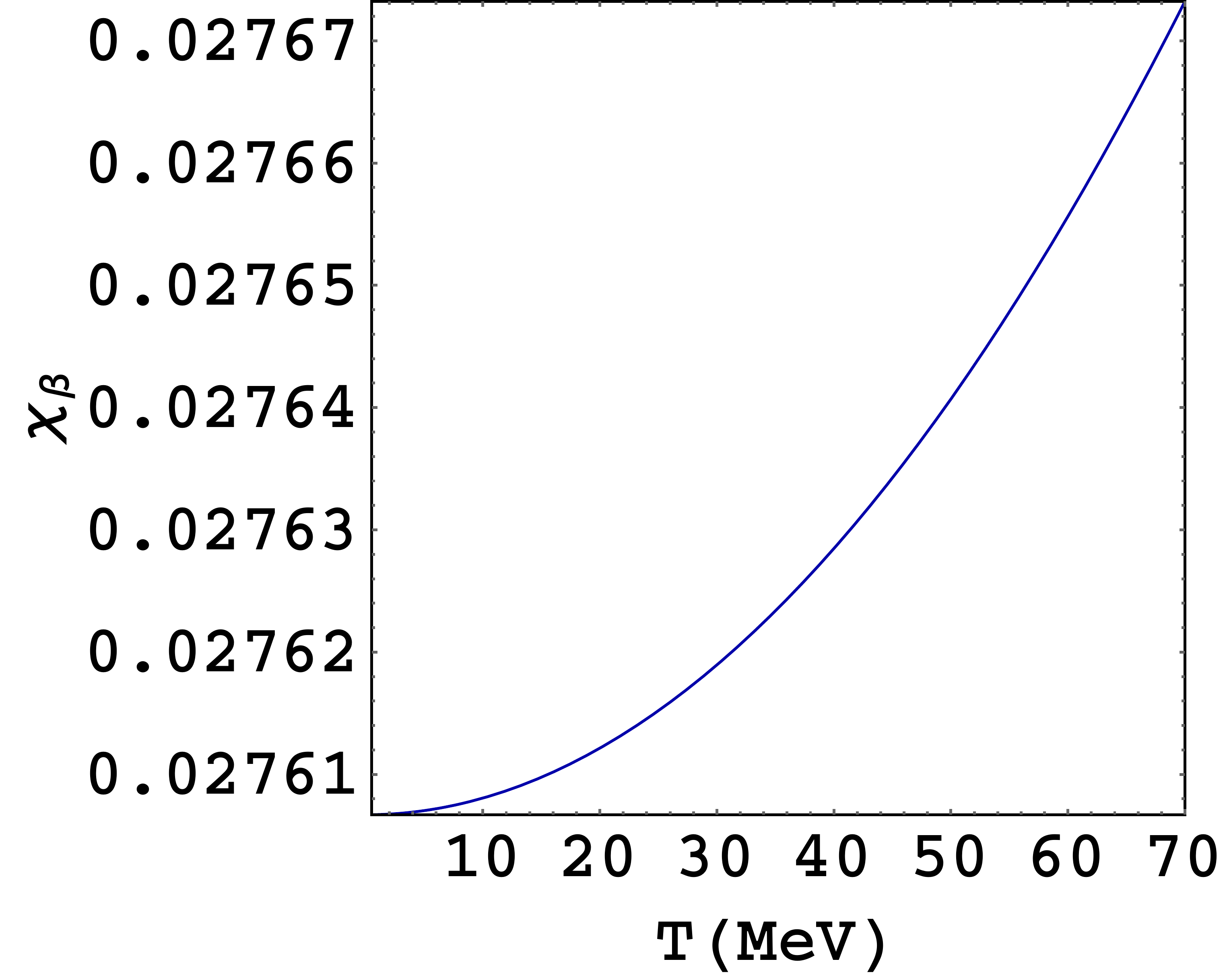}
\caption{(Color online) Magnetic susceptibility vs temperature for a fixed baryonic chemical potential $\mu=1.1$ GeV.}
\label{Sus-T}
\end{center}
\end{figure}

In Fig. \ref{Sus-T}, in order to distinguish the temperature effect in the magnetic susceptibility, we plotted $\chi_\beta=\chi-\chi_{QFT}$, versus temperature. Here, $\chi$ is the entire system susceptibility and $\chi_{QFT}$ is the one obtained from $M_{QFT}=-\partial \Omega_{QFT}/\partial B$. We notice that the magnetic susceptibility increases with the temperature in the range of temperatures of interest for proto-NSs. The physical explanation for this peculiar behavior can be found by taking into account that the temperature acts by pumping out particles from the Fermi sphere.
At relatively low temperatures, the particles remain mostly confined inside the Fermi sphere, where the total magnetic moment in the deep states is always zero because of the filling of the energy levels in agreement with Pauli principle (i.e., a spin-up state will be always accompanied by a spin-down state). The temperature releases  this constraint by evaporating the Fermi sphere, which gives rise to the possibility of more states with the same magnetic moment orientation.

On the other hand, in the case that the particle number density is fixed instead of the baryonic chemical potential, it can be shown that $\chi_\beta \approx 1/T$, which is the typical behavior of paramagnetism, known as Curie's law.

\section{Magnetic-field dependent Specific Heat }\label{section4}

 The role of thermal effects in the presence of a magnetic field is of major importance for understanding the transport properties of NSs from which the thermal evolution of these compact objects can be determined. With the recent possibility of studying binary NS mergers through the entire electromagnetic spectrum \cite{EM-Spectrum} and/or by gravitational waves \cite{Merger}, a new window of opportunity has opened to search for the inner properties of postmerger objects, which could help discriminate between the different EOS candidates. In this direction, having an understanding of NS thermal evolution could be crucial  \cite{Alford}.
 
 In this sense, the specific heat $C_V$ of the NS medium becomes a central quantity \cite{Weber}. As is known, the specific heat of NS is dominated by neutrons, which are the particles in this medium with the largest phase space of low-energy excitations. In the cooling process of a star, the specific heat is proportional to the time needed to reach thermal equilibrium. This is easily seen from the energy balance equation in the Newtonian formulation
  \begin{equation}
	\frac{dE_{th}}{dt}=C_V\frac{dT}{dt},
\end{equation}\label{energy-balance}
where $E_{th}$ is the thermal energy content of the star and $T$ is the internal temperature. 
Thus, to know how much an applied magnetic field can affect the $C_V$ of neutrons it is important to determine the star's capacity to conduct away a significant amount of extra heat.
  
 The specific heat can be determined from the thermodynamic potential through the relationship
 
\begin{equation}
    C_{V}=-T\frac{\partial^2\Omega_S}{\partial{T}^2},
\end{equation}

From (\ref{Omega-N})-(\ref{Omega-beta}), we have
\begin{equation}\label{Cv}
   C_{V} =\frac{1}{32\pi^3T^2}\sum_\sigma\int{d^3}p\Big[(E_{+,\sigma}+\mu)^2\sech^2\Big[\frac{E_{+,\sigma}+\mu}{2T}\Big]+(E_{+,\sigma}-\mu)^2\sech^2\Big[\frac{E_{+,\sigma}-\mu}{2T}\Big]\Big].
\end{equation}

\begin{figure}
    \centering
    \includegraphics[scale=.20]{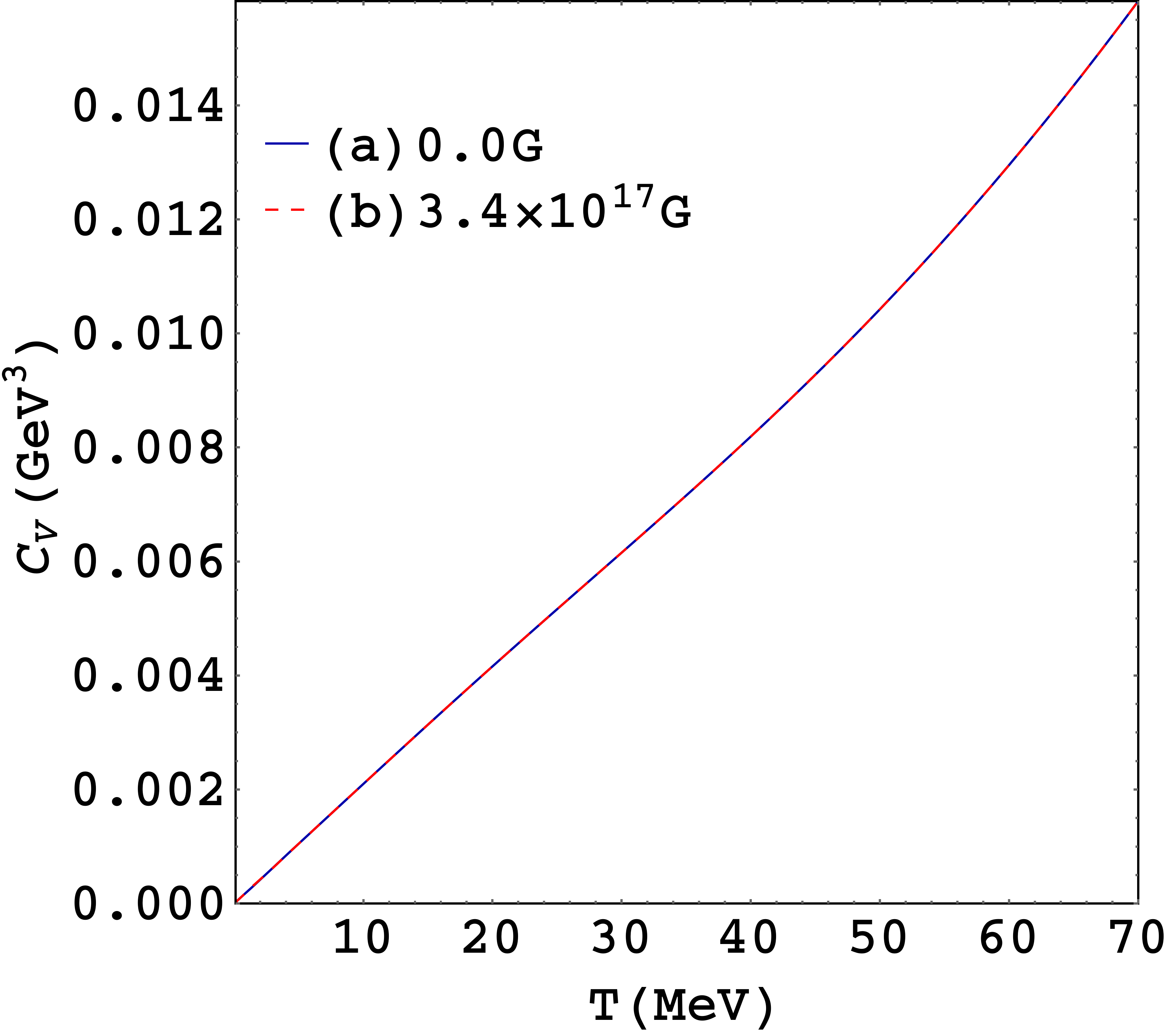}
    \caption{Specific heat at a constant magnetic field strength of (a) $B=0.0G$ (continuous) and (b) $B=3.4\times10^{17}G$ (dashed) as a function of temperature. Note the plots presented here are indistinguishable.}
    \label{CvvT}
\end{figure}

In Fig. \ref{CvvT}, $C_V$ increases monotonically with temperature from $T=0$ MeV to $T=70$ MeV. We also note that up to magnetic fields of $10^{18}$ G, the magnetic-field effect on $C_V$ is negligibly small.

\section{Anisotropic Hydrodynamical equilibrium and estimation of the Neutron Star maximum inner magnetic field } \label{Inner-field}

While the surface magnetic fields of NSs are directly derived from observations, the inner magnetic field can be only inferred from theoretical analysis. The estimated value that is commonly considered is obtained within a Newtonian approach from the equipartition between the gravitational and magnetic energies of a spherical star with homogenous field and mass distributions \cite{Nuclear-matter-field} 
\begin{equation}\label{Uniform-Field}
\left(\frac{4}{3}\pi R^3\right)\frac{B^2}{8\pi}=\frac{3}{5}G\frac{M^2}{R}.
\end{equation}
From Eq. (\ref{Uniform-Field}), it was found that the inner field, $B_0$, is given by
\begin{equation}\label{Max-Uniform-Field}
B_{0}=B_\odot \left(\frac{M}{M_\odot}\right) \left(\frac{R}{R_\odot}\right)^{-2}.
\end{equation}
Using the solar magnetic field value $B_\odot =2\times 10^8$ $G$, and taking into account that a typical neutron star has $M\simeq1.4 M_\odot$ and $R\simeq (0.14\times10^{-4}) R_\odot$, we found that $B_{0}\approx10^{18}$ $G$.

Nevertheless, this value can be challenged by the fact that a sufficiently high magnetic field can provoke a prompt decay of the matter pressure along the field direction, making it impossible to balance the gravitational pull. 
Thus, in this section we want to consider the anisotropic magneto-hydrostatic equilibrium conditions, which take into account the parallel pressure decay, to estimate the maximum value that will be allowed for the inner magnetic field under these conditions. We will keep the Newtonian framework followed in Ref. \cite{Nuclear-matter-field} as a way to get an insight into this problem. Although in a more realistic analysis it should be taken into account that NS are so compact that a general relativistic approach will be necessary. Needless to say, such a treatment will have a high degree of difficulty, since it should consider a metric which is not spherical but in agreement with the cylindrical symmetry imposed by the uniform magnetic field.

Taking into account that the star should be deformed by the anisotropic field effect, we consider that the star shape is ellipsoidal with a smaller radial distance in the direction parallel to the field, which is the one with the lower pressure. 
We note that long ago Chandrasekhar and Fermi, studying the problem of gravitational stability in the presence of a magnetic field, recognized the flattening effect of the magnetic field  \cite{Fermi}.

The surface magnetic field strength will be taken as $10^{14}G$, which is of the order of those found in magnetars. Since the magnetic field is taken to be in the $\hat{z}$ direction, the system should be symmetric about the $z$ axis and the equation for the surface of the star in cylindrical coordinates is given by
\begin{equation}
    \frac{r^{2}}{R^{2}}+\frac{z^{2}}{L^{2}}=1,
\end{equation}
with $L$ being the maximum distance in the $z$ direction and $R=r_{max}$ being the maximum distance in the perpendicular direction to the $z$ axis. 

For stable stars, the matter pressure at any point in the interior of the star is balanced by the gravitational pressure along with any other pressure arising from other sources. In terms of components parallel and perpendicular to the $z$ axis, the pressure balance can be expressed as
\begin{equation} \label{Hydro-Equi-Eq}
    P_{\parallel}(B,\mu)=P_{g_{\parallel}}(r,\phi,z)+P_{other_{\parallel}},\quad P_{\perp}(B,\mu)=P_{g_{\perp}}(r,\phi,z)+P_{other_{\perp}},
\end{equation}
where $ P_{\parallel}$ and $P_{\perp}$, are the parallel and perpendicular matter pressures respectively and $P_{g_{\parallel}}$ and $P_{g_{\perp}}$ are the corresponding gravitational pressures. The $P_{other}$ pressures, as will be introduced in a moment, are due to boundary effects produced by the stellar crust.

The parallel gravitational pressure at a height $z$ along the $z$ axis and the perpendicular gravitational pressure at a distance $r$ along the radial direction with $z=0$ and azimuthal angle $\phi=0$ (i.e., along the direction of $r_{max}$) are approximated as the weight of the column of neutron matter above points $(0,0,z)$ and $(r,0,0)$ respectively, which are given by 
\begin{equation}
    P_{g_{\parallel}}(0,0,z)=-\int_{z}^{L}{\rho(0,0,z')}{\vec{g}(0,0,z'){\cdot}\hat{z}'dz'}, \quad P_{g_{\perp}}(r,0,0)=-\int_{r}^{R}{\rho(r',0,0)}{\vec{g}(r',0,0){\cdot}\hat{r}'dr'}
\end{equation}
where $\hat{z}'$ and $\hat{r}'$ are unit vectors.
Using $\vec{g}(r',\phi',z')=-\nabla{\Phi(r',\phi',z')}$, this becomes
\begin{equation}
    P_{g_{\parallel}}(0,0,z)=\int_{z}^{L}{{\rho(0,0,z')}{\nabla{\Phi(0,0,z')}}{\cdot}\hat{z'}dz'}, \quad P_{g_{\perp}}(r,0,0)=\int_{r}^{R}{{\rho(r',0,0)}{\nabla{\Phi(r',0,0)}}{\cdot}\hat{r'}dr'}.
\end{equation}
    
The gravitational potential as a function of the star's mass density $\rho(\vec{\scriptr}')$ is given as
\begin{equation}
    \Phi(\vec{\scriptr})=-\int_V \frac{{\rho(\vec{\scriptr}')}dV'}{|\vec{\scriptr}-\vec{\scriptr}'|}.
\end{equation}

In cylindrical coordinates it can be written as follows
\begin{equation}
    \Phi(r,\phi,z)=-\int_{0}^{2{\pi}}\int_{0}^{R}\int_{-L\sqrt{1-\frac{{r'}^2}{R^2}}}^{L\sqrt{1-\frac{{r'}^2}{R^2}}}{\frac{{\rho(r',\phi',z')}r'}{\sqrt{(r-r'\cos\phi')^2+(r'\sin\phi')^2+(z-z')^2}}}\,dz'\,dr'\,d\phi'
\end{equation}

Since the gravitational pressure arising from neutron matter vanishes at the surface, while the matter pressure, including that coming from the pure magnetic field, does not, it is natural to find the extra balance in the additional pressures $P_{other_{\parallel/\perp}}$, which we posit is produced by the crust. The pressure arising from the crust along the $z$ and $r$ axes can then be thought of as adding to the gravitational pressures along those directions. We should take into account that the solid crust of a NS makes up only about $1\%$ of the star's total mass, but it will play a basic role here to make the magneto-hydrostatic equilibrium at the surface in the presence of a magnetic field possible.
 
For a star with surface field strength and chemical potential of $B=10^{14}G$ and $\mu=1.1$ GeV respectively, the surface crust pressures are found from 
\begin{equation} 
 P_{crust_{\parallel}}(0,0,L)=P_{\parallel}(10^{14}G,1.1\textrm{GeV}),\quad  P_{crust_{\perp}}(R,0,0)=P_{\perp}(10^{14}G,1.1\textrm{GeV}).
\end{equation}

The solution of the anisotropic magnetohydrostatic equilibrium equations (\ref{Hydro-Equi-Eq}) can give $\mu$ and $B$ at the point $(r=0, z=0)$, once the star geometry $(R,L)$ is fixed.
Then, $B$ and $\mu$ at the star's center can be determined from  
\begin{equation}
    P_{\parallel}(B(0,0,0),\mu(0,0,0))=P_{g_{\parallel}}(0,0,0)+P_{crust_{\parallel}}(0,0,L),
\end{equation}
\begin{equation}
    P_{\perp}(B(0,0,0),\mu(0,0,0))=P_{g_{\perp}}(0,0,0)+P_{crust_{\perp}}(R,0,0).
\end{equation}

From this system of equations, $B(0,0,0)$ and $\mu(0,0,0)$ can be found numerically. For a star with uniform mass density $M=1.4M_\odot$ and dimensions $R=12$ km, $L=10$ km, we find that $B(0,0,0)\approx3.4\times10^{17}$G and $\mu(0,0,0)\approx1.14$ GeV. Hence, we note that in this approach the maximum value of the inner field is an order of magnitude smaller than that found by using the scalar virial theorem in Ref. \cite{Nuclear-matter-field}. That is, the effect of the anisotropy in the magnetohydrostatic equilibrium equations constrains the maximum value  of the field to a lower value. 

Nevertheless, we notice that this maximum value was obtained for a NS with particular values of $R$ and $L$. As follows, we want to relax this condition and introduce a variable eccentricity defined as $e=\sqrt{1-L^2/R^2}$, and see if for other moderate values of $e$ the maximum inner value of $B$ can be larger, in order to reach the $10^{18}$ G virial value.

\begin{figure}
\begin{center}  
    \includegraphics[width=0.47\textwidth]{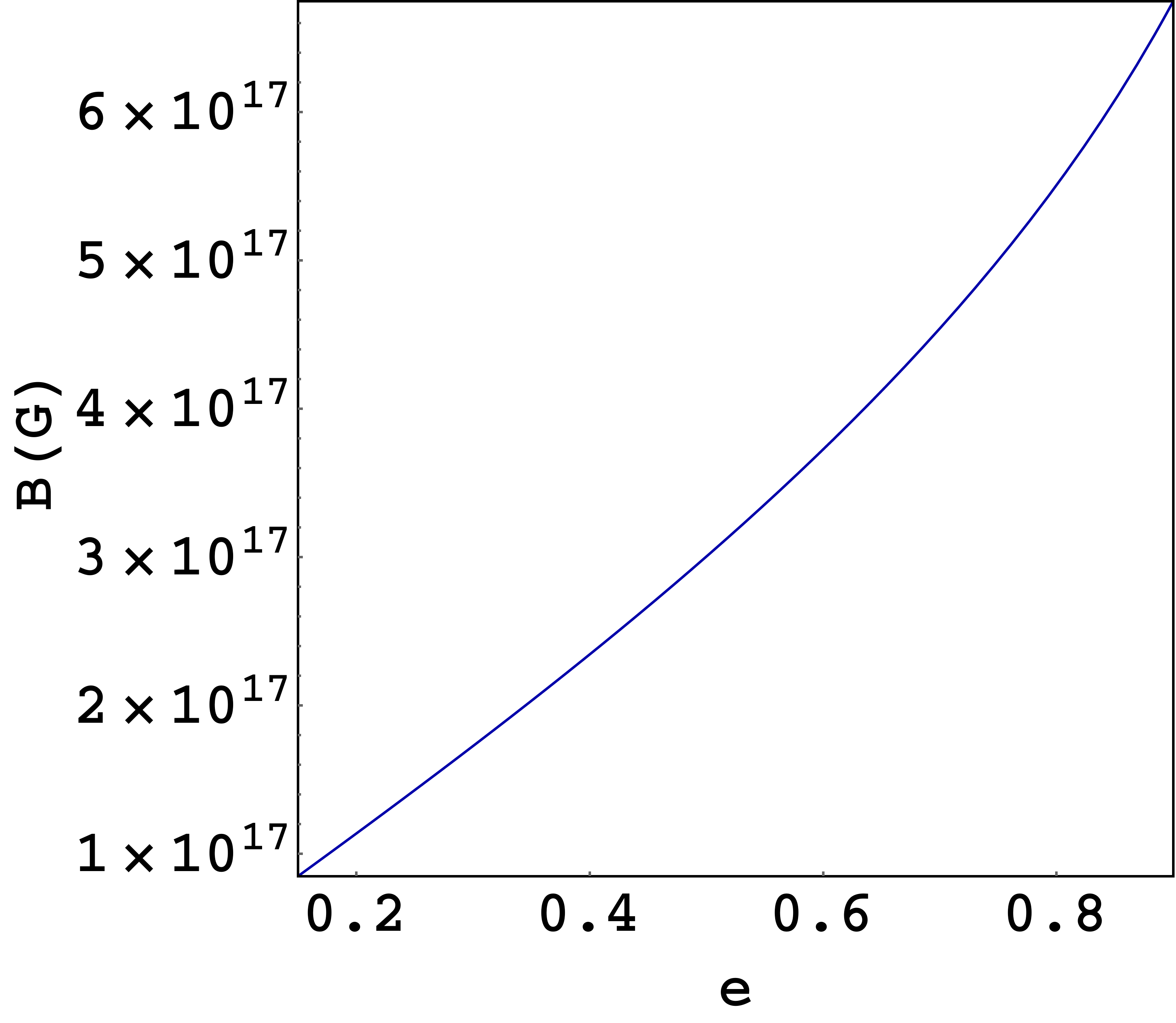}
    \includegraphics[width=0.47\textwidth]{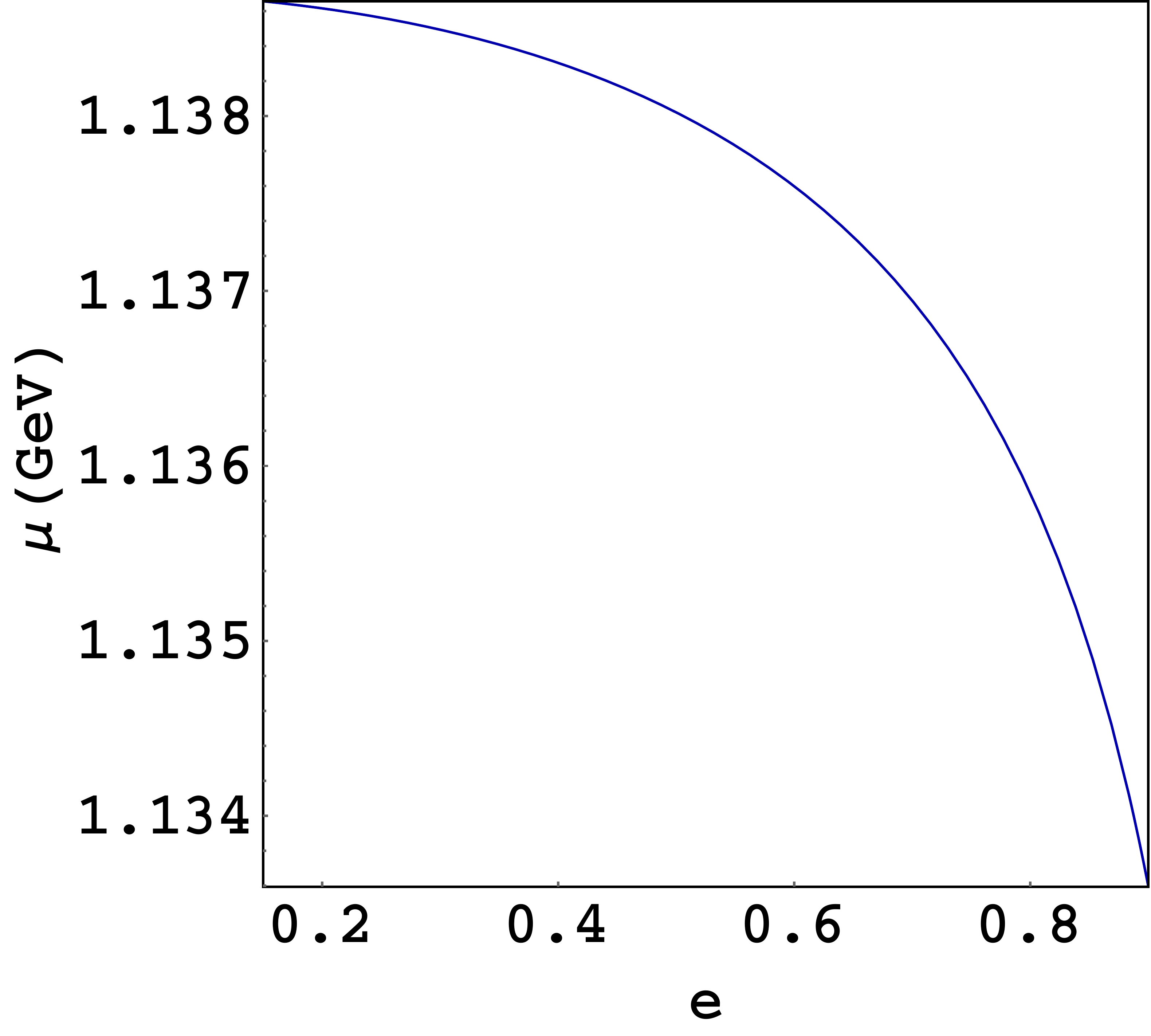}
    \caption{(Left panel) The internal magnetic field strength of an ellipsoidal star at a fixed mass $M=1.4M_\odot$ and volume $V=\frac{4}{3}\pi10^3$ km$^3$ as a function of eccentricity. (Right panel) The internal chemical potential of an ellipsoidal star at a fixed mass $M=1.4M_\odot$ and volume $V=\frac{4}{3}\pi10^3$ km$^3$ as a function of eccentricity.}
    \label{e}
\end{center}
\end{figure}

In Fig. \ref{e}, the inner magnetic field $B(0,0,0)$ is plotted versus the eccentricity of an ellipsoidal star with mass $M=1.4M_\odot$, surface field strength $B=10^{14}$ G, and volume equivalent to a spherical star of radius $R=10$ km. The corresponding central magnetic field strength and chemical potential are presented respectively in the left and right panels. As can be inferred from both panels, the central magnetic field strength increases with eccentricity while maintaining a chemical potential at approximately $\mu=1.1$ GeV. Central field strengths that are close to the parallel pressure  (with Maxwell contribution included) zero value ($B \approx1.5\times10^{18}$G) at $\mu=1.1$ GeV are only reached for large, possibly nonphysical eccentricities. This, along with the analysis done in Sec. \Romannum{3}, suggests that magnetic fields that are large enough to significantly affect the shifting of the EOS are not attainable in a neutron star. 

Here, the following comment is in order. In Ref. \cite{Cardall}, it was also found that the maximum strength of the poloidal inner field in NS is of order $10^{17}$ G, as higher fields resulted in the formation of a torus.

\section{Concluding remarks}\label{section8}

In this paper, we studied how a constant and uniform magnetic field can affect the thermodynamics of a neutron system, characterized by a baryonic chemical potential $\mu$ and a finite temperature $T$. 
The results of this paper are of interest for understanding the physics of a magnetized many-particle neutron system in general, but in particular, our study is mainly motivated by the possibility of having extremely strong magnetic fields in a class of neutron stars known as magnetars. Knowing how a strong magnetic field can alter the physics of such a system is fundamental for determining several important properties of these compact objects, such as their mass-radius relationship, transport properties in their first lifecycle after their formation as a product of a supernova collapse or of a neutron star merger, etc. Although the real composition of a NS is richer than that of a simple many-particle neutron system and should include other baryons, mesons, and leptons, the understanding of the effects of a magnetic field on a pure neutron system is an essential component of the whole picture. Since neutrons form a substantial component of this medium, and because they do not suffer from the softening of pressure due to the Landau momentum quatization, they have the possibility to significantly contribute to the system thermodynamics. 

The first thing to be observed is that in order to have a well-defined Fermi sphere for a neutron many-particle system the condition $\mu^2\gtrsim (M_N+\kappa_NB)^2$ should be satisfied [see Eq. (\ref{Omega-mu-2})]. Taking into account that $M_N=939.56$ MeV, the previous constraint implies that $\mu \gtrsim 1$ GeV. On the other hand, when the magnetized vacuum contribution to the pressure, given by $-\Omega_{QFT}$, is included, we found that a stiffening in the system EOS
takes place.
 
In this scenario, where we are considering chemical potentials up to $2.1$ GeV, the possibility of an interior composition having hyperons becomes highly likely. The reason for this is that at densities of order (2 $-$ 3$)\rho_s$ (which corresponds to chemical potentials $\mu \approx 1.2$ GeV), hyperons are stable against the possible decay into nucleons through weak interactions \cite{Hyperons}. Thus, at the baryon chemical potential required to form the needed Fermi sphere with a corresponding positive pressure at moderate magnetic fields, the conversion of nucleons into hyperons becomes energetically favorable. On the other hand, as is known, the presence of hyperons releases the Fermi pressure exerted by the nucleons and makes the EOS soft enough to lead to a significant reduction of the star mass \cite{hyperons-star-mass}. Hence, as a result of this investigation, we found that a magnetic field is negatively affecting the possibility to reach the 2$M_{\odot}$ with a nucleon inner phase.

Another outcome of this investigation is that due to the pressure anisotropy in the presence of a magnetic field, the magnetohydrostatic equilibrium between the matter and gravitational pressures reduces the inner maximum value that the magnetic field can reach in a compact star when compared with the value that was obtained by using the scalar virial theorem for nuclear matter in Ref. \cite{Nuclear-matter-field}. Now the maximum field for a $1.4 M_\odot$, $R=12$ km star is an order of magnitude smaller than the $10^{18}$ G reported in Ref. \cite{Nuclear-matter-field}. At the new value of order $10^{17}$ G, even if we do not take into consideration the $-\Omega_{QFT}$ pressure, as has been the case in all previous works, the magnetic-field effect on the NS matter EOS is negligible, as was the case with the B-AMM effects on charged fermions \cite{Insignificance}.

But, even when the $-\Omega_{QFT}$ pressure is taken into account, we have shown in Fig. \ref{EOS-Maxwell} that the magnetic field, up to the allowed inner value of $B(0,0,0)\approx3.4\times10^{17}$ G, has an insignificant effect in the system EOS at the required baryon chemical potential $\mu = 1.14$ GeV.
These results contradict the belief accepted at present that sufficiently strong  inner magnetic fields
can produce significant stiffening of the EOS of a neutron system \cite{Lattimer}. On the other hand, under the isotropic consideration and neglecting the QFT contribution, only for $B \approx 10^{18}$ G  does the EOS stiffen significantly \cite{Lattimer}. In our anisotropic approach this field value is beyond the one allowed by the magnetohydrostatic equilibrium condition. 

Finally, we study some thermal effects of this magnetized system. We found that in the temperature region of interest for NS, 0-70 MeV, the system's magnetic susceptibility increases with the temperature.
This behavior is physically explained by taking into account the thermal creation at a fixed $\mu$ of more particles that can contribute to the system magnetization, as explained in Sec. \Romannum{4} . The other thermodynamic quantity we studied was the specific heat of the magnetized neutron system. We found that for the temperature range of interest for proto-NS the magnetic-field effect on $C_V$ is negligibly small.

We should mention that there exists an alternative approach for formulating the magnetic-moment and magnetic-field interaction for  composed particles in a magnetic field. It is based on a Klein-Gordon Lagrangian with a full Pauli dipole moment term \cite{Jan}. The two formulations disagree in their contributions to the Lamb shift  and with respect to the  fine  structure  splitting.  It would be interesting to investigate if they also disagree regarding the magnetic field effects on the system thermodynamics. If a significant disagreement is found in this contest, it could help in determining which approach is better aligned with NS observations.  

It will be of interest to investigate how the anisotropy in the pressures and the consideration of the QFT contribution in the energy density and pressures can affect the EOS of magnetized NS with a more realistic composition that includes other baryons, mesons, and leptons.

\acknowledgments
This work was supported in part by NSF grant PHY-1714183. We thank J. L. Lattimer for some useful comments.

\end{document}